arXiv:1609.07590v1 [quant-ph] 24 Sep 2016

# A Numerical Approach to Optimal Coherent Quantum LQG Controller Design Using Gradient Descent ⋆

Arash Kh. Sichani [a], Igor G. Vladimirov [a], Ian R. Petersen [a]

[a] *School of Engineering and Information Technology, University of New South Wales Canberra, ACT 2600, Australia*

**Abstract**

This paper is concerned with coherent quantum linear quadratic Gaussian (CQLQG) control. The problem is to find a stabilizing measurement-free quantum controller for a quantum plant so as to minimize a mean square cost for the fully quantum closed-loop system. The plant and controller are open quantum systems interconnected through bosonic quantum fields. In comparison with the observation-actuation structure of classical controllers, coherent quantum feedback is less invasive to the quantum dynamics. The plant and controller variables satisfy the canonical commutation relations (CCRs) of a quantum harmonic oscillator and are governed by linear quantum stochastic differential equations (QSDEs). In order to correspond to such oscillators, these QSDEs must satisfy physical realizability (PR) conditions in the form of quadratic constraints on the state-space matrices, reflecting the CCR preservation in time. The symmetry of the problem is taken into account by introducing equivalence classes of coherent quantum controllers generated by symplectic similarity transformations. We discuss a modified gradient flow, which is concerned with norm-balanced realizations of controllers. A line-search gradient descent algorithm with adaptive stepsize selection is proposed for the numerical solution of the CQLQG control problem. The algorithm finds a local minimum of the LQG cost over the parameters of the Hamiltonian and coupling operators of a stabilizing coherent quantum controller, thus taking the PR constraints into account. A convergence analysis of the algorithm is presented. Numerical examples of designing locally optimal CQLQG controllers are provided in order to demonstrate the algorithm performance.

## 1 Introduction

Coherent quantum feedback control [14,16] is a quantum control paradigm which is aimed at achieving given performance specifications for quantum systems, such as internal stability and optimization of a cost functional. Such systems arise naturally in quantum physics [9] and its engineering applications (for example, nanotechnology and quantum optics [6]). The dynamic variables of quantum systems are (usually noncommuting) operators on an underlying Hilbert space which evolve according to the laws of quantum mechanics [18]. The quantum dynamics are particularly sensitive to interaction with classical devices over the course of quantum measurement, as reflected in the projection postulate of quantum mechanics. In order to overcome this issue, coherent quantum control employs the idea of direct interconnection of quantum plants (that is, the quantum systems to be controlled) with other quantum systems playing the role of controllers, possibly mediated by light fields. Unlike the traditional observation-actuation control loop, this fully quantum measurement-free feedback avoids the loss of quantum information resulting from a conversion to classical signals.

Quantum-optical components, such as optical cavities, beam splitters and phase shifters, make it possible to implement coherent quantum feedback governed by Markovian linear quantum stochastic differential equations (QSDEs) [23,24], provided the latter are physically realizable (PR) as open quantum harmonic oscillators [5,6]. The resulting PR conditions [11,25,26] are organized as quadratic constraints on the coefficients of the QSDEs. The PR constraints for the state-space matrices of a

 Corresponding author: Arash Kh. Sichani. Tel. +61-2-626-89462. Fax +61-2-626-88443.

*Email addresses:* arash_kho@hotmail.com (Arash Kh. Sichani), igor.g.vladimirov@gmail.com (Igor G. Vladimirov), i.r.petersen@gmail.com (Ian R. Petersen).

coherent quantum controller complicate the solution of quantum counterparts to the classical Linear Quadratic Gaussian (LQG) and $\mathscr{H}_\infty$ control problems.

The Coherent Quantum LQG (CQLQG) control problem [21] seeks for a stabilizing PR quantum controller so as to minimize an infinite-horizon mean square cost functional for the fully quantum closed-loop system. This is a constrained optimization problem for the steady-state quantum covariance matrix of the plant-controller system satisfying an algebraic Lyapunov equation (ALE). A numerical procedure for finding *suboptimal* controllers for this problem was proposed in [21], and algebraic equations for the *optimal* CQLQG controller were obtained in [33]. We also mention that coherent quantum LQG control settings were considered in [15] for a class of quantum systems (with annihilation operators only), in the context of evolutionary optimization for entanglement control [7], and also for different scenarios of plant-controller coupling in [35]. Despite the previous results, the CQLQG control problem does not lend itself to an "elegant" solution (for example, in the form of decoupled Riccati equations as in the classical case [12]) and remains a subject of research. Since the main difficulties are caused by the coupling between the ALEs for the state-space matrices of the optimal controller due to the PR constraints, [34] offered an alternative approach based on moving the "burden" of the constraints to the Lagrange multipliers for a coherent quantum filtering problem [19] which is a simplified feedback-free version of the control problem.

In the present paper, we develop an algorithm for the numerical solution of the CQLQG control problem by using a line-search (gradient descent) method and the Hamiltonian parameterization of PR quantum controllers [33]. This parameterization is a different technique to handle the PR constraints by reformulating the CQLQG control problem in an unconstrained fashion. More precisely, the optimal solution is sought in the class of stabilizing PR controllers whose state-space matrices are parameterized in terms of the free Hamiltonian and coupling operators of an open quantum harmonic oscillator [5]. We obtain ordinary differential equations (ODEs) for the gradient descent in the Hilbert space of these matrix-valued parameters of coherent quantum controllers. In accordance with the PR conditions, the CQLQG control problem has a special type of *symmetry* which makes it invariant under symplectic similarity transformations of the controller variables [31,33]. We take this symmetry into account and consider equivalence classes of state-space representations of coherent quantum controllers. We also propose a modified gradient flow, which is concerned with norm-balanced realizations of such controllers and resembles the steepest descent with respect to a different Riemannian metric [2]. For this purpose, we combine the Fréchet and Gâteaux differentiation with differential geometric tools (such as Lie groups [22] and tangent spaces) and related algebraic techniques [3,17,29,32,33] to employ the analytic structure of the LQG cost as a composite function of the matrix-valued variables, whose computation involves ALEs.

A useful feature of the gradient descent approach to the CQLQG control problem is that, at intermediate steps, it produces stabilizing PR quantum controllers which can be regarded as gradually improving suboptimal solutions of the problem, and a locally optimal solution (if it exists) is achieved asymptotically by moving along anti-gradient directions with a suitable choice of stepsizes. To this end, we provide an algorithm for adaptive stepsize selection for each iteration based on the second-order Gâteaux (directional) derivative of the LQG cost along the gradient. However, the proposed gradient descent algorithm for the CQLQG control problem requires for its initialization a *stabilizing* PR quantum controller. Finding such a controller for an arbitrary given quantum plant is a nontrivial open problem which has recently been considered in the frequency domain [27]. Because of the lack of a systematic solution for this *quantum stabilization problem*, the present algorithm is initialized at a stabilizing PR quantum controller which is obtained by a random search in the space defined by the Hamiltonian parameterization of PR controllers. Although a random search of an admissible starting point is acceptable for low-dimensional problems, the development of a more systematic approach to this issue is a subject of future research.

The paper is organized as follows. Section 2 outlines the principal notation. Sections 3 and 4 specify the quantum plants and coherent quantum controllers being considered. Section 5 revisits the PR conditions for linear quantum systems. Section 6 formulates the CQLQG control problem. Section 7 specifies the gradient flow for finding local minima in this problem. Section 8 defines equivalence classes generated by symplectic similarity transformations of coherent quantum controllers. These are used in Section 9 which is concerned with norm-balanced realizations of coherent quantum controllers and proposes a modified gradient flow. Section 10 describes an algorithmic implementation of the gradient descent method with an adaptive line search. Section 11 discusses convergence of this algorithm. Section 12 provides numerical examples of designing locally optimal CQLQG controllers. Section 13 gives concluding remarks. Appendices A and B provide a subsidiary material on the differentiation of the LQG cost.

## 2 Notation

Vectors are assumed to be organized as columns unless specified otherwise, and the transpose $(\cdot)^{\mathrm{T}}$ acts on matrices with operator-valued entries as if the latter were scalars. For a vector $X$ of operators $X_1, \ldots, X_r$ and a vector $Y$ of operators $Y_1, \ldots, Y_s$, the commutator matrix $[X, Y^{\mathrm{T}}] := XY^{\mathrm{T}} - (YX^{\mathrm{T}})^{\mathrm{T}}$ is an $(r \times s)$-matrix whose $(j,k)$th entry is the commutator $[X_j, Y_k] := X_jY_k -$ $Y_kX_j$ of the operators $X_j$ and $Y_k$. Furthermore, $(\cdot)^\dagger := ((\cdot)^\#)^{\mathrm{T}}$ denotes the transpose of the entry-wise operator adjoint $(\cdot)^\#$. When it is applied to complex matrices, $(\cdot)^\dagger$ reduces to the complex conjugate transpose $(\cdot)^* := (\overline{(\cdot)})^{\mathrm{T}}$. Denoted by $\mathrm{sym}(\cdot) := \frac{(\cdot)+(\cdot)^{\mathrm{T}}}{2}$ and $\mathrm{asym}(\cdot) := \frac{(\cdot)-(\cdot)^{\mathrm{T}}}{2}$ are the symmetrizer and antisymmetrizer of matrices. Also, we denote by $\mathbb{S}_r$, $\mathbb{A}_r$ and $\mathbb{H}_r := \mathbb{S}_r + i\mathbb{A}_r$ the subspaces of real symmetric, real antisymmetric and complex Hermitian matrices of order $r$, respectively, with $i := \sqrt{-1}$ the imaginary unit. Denoted by $\mathbf{J} := \begin{bmatrix} 0 & 1 \\ -1 & 0 \end{bmatrix}$ is a matrix which spans the space $\mathbb{A}_2$. Furthermore, $I_r$ denotes the identity matrix of order $r$, positive (semi-) definiteness of matrices is denoted by ($\succcurlyeq$) $\succ$, and $\otimes$ is the tensor product of spaces or operators (in particular, the Kronecker product of matrices). The adjoints and self-adjointness of linear operators acting on matrices is understood in the sense of the Frobenius inner product $\langle M, N\rangle := \mathrm{Tr}(M^*N)$ of real or complex matrices, with the corresponding Frobenius norm $\|M\| := \sqrt{\langle M, M\rangle}$ which is the standard Euclidean norm $|\cdot|$ for vectors. Also, $\mathbf{E}X := \mathrm{Tr}(\rho X)$ denotes the quantum expectation of a quantum variable $X$ (or a vector of such variables) over a density operator $\rho$ which specifies the underlying quantum state.

## 3 Quantum Plant

The quantum plant under consideration is an open quantum stochastic system which is coupled to another such system (playing the role of a controller), with the dynamics of both systems being affected by the environment. Both the plant and the controller are assumed to satisfy the physical realizability (PR) conditions [11,21,25,26] which will be described in Section 5. The plant has an even number $n$ of dynamic variables $x_1(t), \ldots, x_n(t)$ which are time-varying self-adjoint operators on a Hilbert space specified below. With the time arguments being omitted for brevity, the evolution of the vector $x := (x_k)_{1\leqslant k\leqslant n}$ of plant variables and its contribution to a $p_1$-dimensional output of the plant $y := (y_k)_{1\leqslant k\leqslant p_1}$ (also with self-adjoint operator-valued entries) are governed by the QSDEs

$$\mathrm{d}x = Ax\mathrm{d}t + B\mathrm{d}w + E\mathrm{d}\eta, \qquad \mathrm{d}y = Cx\mathrm{d}t + D\mathrm{d}w. \tag{1}$$

Here, $A \in \mathbb{R}^{n\times n}$, $B \in \mathbb{R}^{n\times m_1}$, $C \in \mathbb{R}^{p_1\times n}$, $D \in \mathbb{R}^{p_1\times m_1}$, $E \in \mathbb{R}^{n\times p_2}$ are given constant matrices. Also,

$$z := Cx \tag{2}$$

is a "signal part" of the plant output $y$, and $\eta$ is a $p_2$-dimensional output of the controller to be described in Section 4. The external noise acting on the plant is represented by a quantum Wiener process $w := (w_k)_{1\leqslant k\leqslant m_1}$ whose entries are self-adjoint operators on a boson Fock space $\mathscr{F}_1$ [23] with the quantum Itô table $\mathrm{d}w\mathrm{d}w^{\mathrm{T}} = \Omega_1\mathrm{d}t$, where the matrix $\Omega_1 \in \mathbb{H}_{m_1}$ is given by $\Omega_1 := I_{m_1} + iJ_1 \succcurlyeq 0$. Here, the matrix $J_1 \in \mathbb{A}_{m_1}$ specifies the CCRs between the entries of the plant noise $w$ as $[\mathrm{d}w, \mathrm{d}w^{\mathrm{T}}] = 2iJ_1\mathrm{d}t$ and (assuming that the dimension $m_1$ is even) is given by $J_1 := I_{m_1/2} \otimes \mathbf{J}$.

## 4 Quantum Controller

Consider an interconnection of the plant (1) with a coherent (that is, measurement-free) quantum controller. The latter is also an open quantum system with an $n$-dimensional state vector $\xi := (\xi_k)_{1\leqslant k\leqslant n}$ of self-adjoint operators on another Hilbert space, which also evolve in time. The assumption that the controller has the same number of dynamic variables as the plant is adopted from the classical LQG control theory. The controller interacts with the plant and the environment according to the QSDEs

$$\mathrm{d}\xi = a\xi\mathrm{d}t + b\mathrm{d}\omega + e\mathrm{d}y, \qquad \mathrm{d}\eta = c\xi\mathrm{d}t + d\mathrm{d}\omega. \tag{3}$$

Here, $a \in \mathbb{R}^{n\times n}$, $b \in \mathbb{R}^{n\times m_2}$, $c \in \mathbb{R}^{p_2\times n}$, $d \in \mathbb{R}^{p_2\times m_2}$, $e \in \mathbb{R}^{n\times p_1}$ are also constant matrices. Similarly to (2), the $p_2$-dimensional process

$$\zeta := c\xi \tag{4}$$

is the signal part of the controller output $\eta$. The process $\omega$ in (3) is a quantum noise which affects the controller and is an $m_2$-dimensional quantum Wiener process (with $m_2$ even) on another boson Fock space $\mathscr{F}_2$ with the quantum Ito table $\mathrm{d}\omega\mathrm{d}\omega^{\mathrm{T}} = \Omega_2\mathrm{d}t$, where the matrix $\Omega_2 \in \mathbb{H}_{m_2}$ is given by $\Omega_2 := I_{m_2} + iJ_2 \succcurlyeq 0$. Here, the matrix $J_2 \in \mathbb{A}_{m_2}$ specifies the CCRs between the entries of the controller noise $\omega$ as $[\mathrm{d}\omega, \mathrm{d}\omega^{\mathrm{T}}] = 2iJ_2\mathrm{d}t$ and is given by $J_2 := I_{m_2/2} \otimes \mathbf{J}$. The plant and controller noises $w$ and $\omega$ act on different boson Fock spaces $\mathscr{F}_1$ and $\mathscr{F}_2$, respectively, and hence, commute with each other. Therefore, the combined quantum Wiener process

$$\mathscr{W} := \begin{bmatrix} w \\ \omega \end{bmatrix} \tag{5}$$

of dimension $m := m_1 + m_2$ acts on the tensor product space $\mathscr{F}_1 \otimes \mathscr{F}_2$ and has a block diagonal CCR matrix $J$:

$$[\mathrm{d}\mathscr{W}, \mathrm{d}\mathscr{W}^{\mathrm{T}}] = 2iJ\mathrm{d}t, \qquad J := \begin{bmatrix} J_1 & 0 \\ 0 & J_2 \end{bmatrix}. \tag{6}$$

Furthermore, the external boson fields are assumed to be in the product vacuum state $\upsilon := \upsilon_1 \otimes \upsilon_2$, and hence, are uncorrelated. The resulting quantum Ito table of the combined Wiener process $\mathscr{W}$ in (5) is

$$\mathrm{d}\mathscr{W}\mathrm{d}\mathscr{W}^{\mathrm{T}} = \Omega \mathrm{d}t, \qquad \Omega := I_m + iJ = \Omega^* \succcurlyeq 0. \tag{7}$$

In the controller dynamics (3), the matrix $b$ is the noise gain matrix, while $e$ plays the role of the observation gain matrix, although $y$ is not an observation signal in the classical control theoretic sense. Accordingly, the process $\zeta$ in (4) corresponds to the classical actuator signal. Now, the combined set of QSDEs (1), (3) describes the fully quantum closed-loop system shown in Fig. 1. By using a quadratic cost, adopted in quantum control settings [21,33] from classical LQG control [12], the performance of the coherent quantum controller will be described in Section 6 in terms of an $r$-dimensional quantum process

$$\mathscr{Z} := Fx + G\zeta. \tag{8}$$

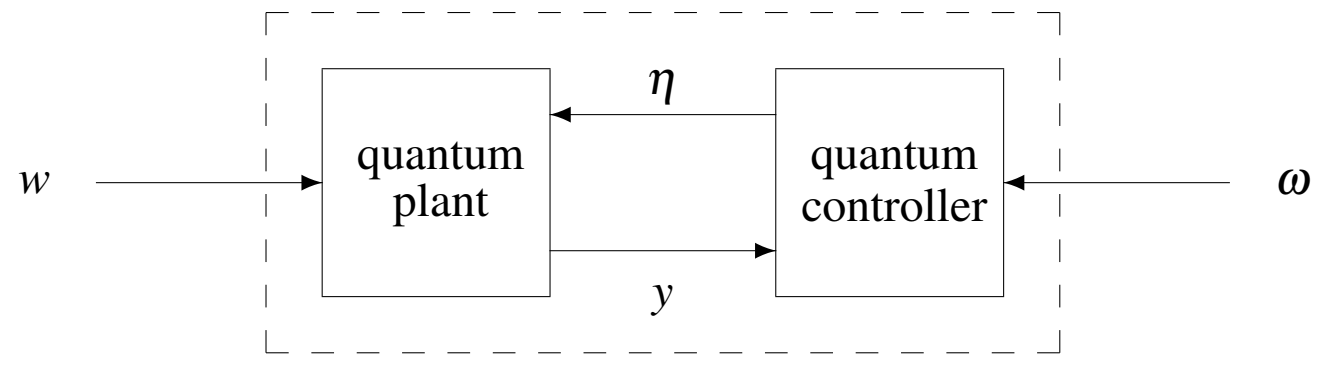

Fig. 1. The interconnected quantum plant and quantum controller form a fully quantum closed-loop system which is governed by (1), (3) and is influenced by the environment through the quantum Wiener processes $w$, $\omega$.

Here, $F \in \mathbb{R}^{r\times n}$, $G \in \mathbb{R}^{r\times p_2}$ are given *weighting matrices* whose entries quantify the relative importance of the state variables $x_1, \ldots, x_n$ of the plant and the "actuator output" variables $\zeta_1, \ldots, \zeta_{p_2}$ of the controller. The choice of $F$, $G$ is specified only by control design preferences and is not subjected to physical constraints. The process $\mathscr{Z}$ in (8) is linearly related to the $2n$-dimensional vector of dynamic variables

$$\mathscr{X} := \begin{bmatrix} x \\ \xi \end{bmatrix} \tag{9}$$

of the closed-loop system whose dynamics are governed by the QSDE

$$\mathrm{d}\mathscr{X} = \mathscr{A}\mathscr{X}\mathrm{d}t + \mathscr{B}\mathrm{d}\mathscr{W}, \qquad \mathscr{Z} = \mathscr{C}\mathscr{X} \tag{10}$$

which is driven by the combined quantum Wiener process $\mathscr{W}$ from (5). The state-space matrices $\mathscr{A} \in \mathbb{R}^{2n\times 2n}$, $\mathscr{B} \in \mathbb{R}^{2n\times m}$, $\mathscr{C} \in \mathbb{R}^{r\times 2n}$ of the closed-loop system in (10) are obtained by combining (1), (3) with (5), (8), (9) and depend on the controller matrices $a$, $b$, $c$, $e$ in an affine fashion:

$$\mathscr{A} := \begin{bmatrix} A & Ec \\ eC & a \end{bmatrix}, \qquad \mathscr{B} := \begin{bmatrix} B & Ed \\ eD & b \end{bmatrix}, \qquad \mathscr{C} := \begin{bmatrix} F & Gc \end{bmatrix}. \tag{11}$$

## 5 Conditions of Physical Realizability

Both the quantum plant (1) and the coherent quantum controller (3) are assumed to be physically realizable as open quantum harmonic oscillators, with initial complex separable Hilbert spaces $\mathscr{H}_1$, $\mathscr{H}_2$. In particular, their dynamic variables (which are self-adjoint operators on the product space $\mathscr{H}_1 \otimes \mathscr{H}_2 \otimes \mathscr{F}_1 \otimes \mathscr{F}_2$ at any subsequent moment of time $t > 0$) satisfy CCRs

$$[x, x^{\mathrm{T}}] = 2i\Theta_1, \qquad [\xi, \xi^{\mathrm{T}}] = 2i\Theta_2, \qquad [x, \xi^{\mathrm{T}}] = 0, \tag{12}$$

where $\Theta_1, \Theta_2 \in \mathbb{A}_n$ are constant nonsingular matrices. An equivalent form of the CCRs for the combined vector $\mathscr{X}$ from (9) is

$$[\mathscr{X}, \mathscr{X}^{\mathrm{T}}] = 2i\Theta, \qquad \Theta := \begin{bmatrix} \Theta_1 & 0 \\ 0 & \Theta_2 \end{bmatrix}. \tag{13}$$

The preservation of the CCRs (12) (including the commutativity between $x$ and $\xi$) is a consequence of the unitary evolution of the isolated system formed from the plant, controller and their environment. The QSDE in (10) preserves the CCR matrix $\Theta$ in (13) in time if and only if the matrices $\mathscr{A}$, $\mathscr{B}$ in (11) satisfy

$$\mathscr{A}\Theta + \Theta\mathscr{A}^{\mathrm{T}} + \mathscr{B}J\mathscr{B}^{\mathrm{T}} = 0, \tag{14}$$

where $J$ is the CCR matrix of the combined quantum Wiener process $\mathscr{W}$ in (5) given by (6). The relation (14) is obtained by taking the imaginary part of the algebraic Lyapunov equation (ALE)

$$\mathscr{A}S + S\mathscr{A}^{\mathrm{T}} + \mathscr{B}\Omega\mathscr{B}^{\mathrm{T}} = 0 \tag{15}$$

(provided $\mathscr{A}$ is Hurwitz) for the steady-state quantum covariance matrix

$$S := \lim_{t\to+\infty} \mathbf{E}(\mathscr{X}(t)\mathscr{X}(t)^{\mathrm{T}}) = P + i\Theta = S^* \succcurlyeq 0, \tag{16}$$

with $\Omega$ the quantum Ito matrix from (7). Here, the quantum expectation $\mathbf{E}(\cdot)$ is taken over the product state $\varpi \otimes \upsilon$, where $\varpi$ is the initial quantum state of the plant and controller on $\mathscr{H}_1 \otimes \mathscr{H}_2$, and $\upsilon$ is the vacuum state of the external fields on $\mathscr{F}_1 \otimes \mathscr{F}_2$. We have also used the convergence $\lim_{t\to+\infty} \mathbf{E}\mathscr{X}(t) = 0$ which is ensured by $\mathscr{A}$ being Hurwitz. The real part

$$P := \mathrm{Re} S \tag{17}$$

of the quantum covariance matrix $S$ from (16) coincides with the controllability Gramian [12] of the pair $(\mathscr{A}, \mathscr{B})$ which is the unique solution of the ALE

$$\mathscr{A}P + P\mathscr{A}^{\mathrm{T}} + \mathscr{B}\mathscr{B}^{\mathrm{T}} = 0, \tag{18}$$

obtained by taking the real part of (15). Since the left-hand side of (14) is an antisymmetric matrix of order $2n$, then, by computing the diagonal $(n \times n)$-blocks and the upper off-diagonal block of this matrix with the aid of (11), it follows that the preservation of the CCR matrix $\Theta$ in (13) is equivalent to

$$A\Theta_1 + \Theta_1 A^{\mathrm{T}} + BJ_1B^{\mathrm{T}} + EdJ_2d^{\mathrm{T}}E^{\mathrm{T}} = 0, \tag{19}$$

$$a\Theta_2 + \Theta_2 a^{\mathrm{T}} + eDJ_1D^{\mathrm{T}}e^{\mathrm{T}} + bJ_2b^{\mathrm{T}} = 0, \tag{20}$$

$$(\Theta_1 C^{\mathrm{T}} + BJ_1D^{\mathrm{T}})e^{\mathrm{T}} + E(c\Theta_2 + dJ_2b^{\mathrm{T}}) = 0; \tag{21}$$

cf. [31, Eqs. (18)–(20)]. Therefore, the fulfilment of the equalities

$$\Theta_1 C^{\mathrm{T}} + BJ_1D^{\mathrm{T}} = 0, \tag{22}$$

$$c\Theta_2 + dJ_2b^{\mathrm{T}} = 0 \tag{23}$$

is sufficient for (21). Note that (19) and (22) are the conditions for physical realizability (PR) [11,21,25,26] of the quantum plant which describe the preservation of the CCR matrix $\Theta_1$ in (12) and $[x, y^{\mathrm{T}}] = 0$. The latter is related to non-demolition conditions for the plant output $y$ with respect to the internal dynamics of the plant. Similarly, the relations (20), (23), which describe the preservation of the CCR matrix $\Theta_2$ in (12) and $[\xi, \eta^{\mathrm{T}}] = 0$, are the PR conditions for the coherent quantum controller. The PR condition (20) can be regarded as a linear equation with respect to the matrix $a$, and its general solution is representable as

$$a = 2\Theta_2 R - \frac{1}{2}(eDJ_1D^{\mathrm{T}}e^{\mathrm{T}} + bJ_2b^{\mathrm{T}})\Theta_2^{-1}. \tag{24}$$

Here, the matrix $R \in \mathbb{S}_n$ specifies the free Hamiltonian $\frac{1}{2}\xi^{\mathrm{T}}R\xi$ which the PR controller would have in the absence of interaction with its surroundings; cf. [5, Eqs. (20)–(22) on pp. 8–9]. The other PR condition (23) allows the matrix $c$ to be expressed in terms of $b$ as

$$c = -dJ_2b^{\mathrm{T}}\Theta_2^{-1}. \tag{25}$$

The coupling between the output matrix $c$ and the noise gain matrix $b$ makes the design of a coherent quantum controller (3) substantially different from that of classical controllers even at the level of achieving internal stability for the closed-loop system. Indeed, if the additional quantum noise $\omega$ is effectively eliminated from the state dynamics of the quantum controller by letting $b = 0$, then (25) implies that $c = 0$, and hence, the matrix $\mathscr{A}$ in (11) becomes block lower triangular. In this case, the closed-loop system in (10) cannot be internally stable if $A$ is not Hurwitz. Also note that, in the formulations of the PR conditions [11,21,25,26] for the plant and controller QSDEs (1), (3), the noise feedthrough matrices are usually given by $D = \begin{bmatrix} I_{p_1} & 0 \end{bmatrix}$ and $d = \begin{bmatrix} I_{p_2} & 0 \end{bmatrix}$, with $p_1 \leqslant m_1$ and $p_2 \leqslant m_2$. Such matrices $D$ and $d$ have full row rank and satisfy

$$DD^{\mathrm{T}} = I_{p_1}, \qquad dd^{\mathrm{T}} = I_{p_2}. \tag{26}$$

The full row rank property of $D$ corresponds to nondegeneracy of measurements in the classical setting, where $y$ in (1) is an observation process. Furthermore, since $\det J_2 \neq 0$ and $\det \Theta_2 \neq 0$, the full row rank property of $d$ implies that the map $\mathbb{R}^{n\times m_2} \ni b \mapsto c \in \mathbb{R}^{p_2\times n}$, given by (25), is onto. This allows the matrix $c$ to be assigned any value by an appropriate choice of $b$, which plays a part in the stabilization issue mentioned above. Although (26) simplifies the algebraic manipulations, it is the rank properties of the matrices $D$, $d$ that are most important.

## 6 The Coherent Quantum LQG Control Problem

Following [21,33], we formulate the CQLQG control problem as that of minimizing the steady-state mean square value

$$\mathscr{E} := \frac{1}{2}\lim_{t\to+\infty} \mathbf{E}(\mathscr{Z}(t)^{\mathrm{T}}\mathscr{Z}(t)) = \frac{1}{2}\langle \mathscr{C}^{\mathrm{T}}\mathscr{C}, P\rangle \longrightarrow \min \tag{27}$$

for the process $\mathscr{Z}$ of the closed-loop system (10) over internally stabilizing (that is, making the matrix $\mathscr{A}$ Hurwitz) PR quantum controllers (3) of fixed dimensions described in Sections 4 and 5. Here, $\mathscr{Z}^{\mathrm{T}}\mathscr{Z} = \sum_{k=1}^{r}\mathscr{Z}_k^2$ is the sum of squared entries of $\mathscr{Z}$ (and hence, $\mathscr{Z}^{\mathrm{T}}\mathscr{Z}$ is a positive semi-definite self-adjoint operator) and $P$ is the controllability Gramian of the closed-loop system given by (17) and (18). The LQG cost $\mathscr{E}$ in (27) is a function of the triple

$$u := (R, b, e) \in \mathbb{S}_n \times \mathbb{R}^{n\times m_2} \times \mathbb{R}^{n\times p_1} =: \mathbb{U} \tag{28}$$

which parameterizes PR quantum controllers (3) through (24) and (25), with the controller noise feedthrough matrix $d \in \mathbb{R}^{p_2\times m_2}$ being fixed and satisfying (26). Accordingly, the minimization in (27) is carried out over the set of those $u$ which specify internally stabilizing PR quantum controllers for the quantum plant (1):

$$\mathbb{U}_0 := \big\{u \in \mathbb{U}:\ \mathscr{A} \text{ in (11) is Hurwitz}\big\}. \tag{29}$$

In what follows, the set $\mathbb{U}$ on the right-hand side of (28) is endowed with the structure of a Hilbert space with the direct-sum inner product

$$\langle (R_1, b_1, e_1), (R_2, b_2, e_2)\rangle := \langle R_1, R_2\rangle + \langle b_1, b_2\rangle + \langle e_1, e_2\rangle \tag{30}$$

and the corresponding norm $\|(R, b, e)\| := \sqrt{\|R\|^2 + \|b\|^2 + \|e\|^2}$. It follows from (28) that the dimension of this space is computed as

$$\dim \mathbb{U} = \dim \mathbb{S}_n + nm_2 + np_1 = n\Big(\frac{1}{2}(n+1) + m_2 + p_1\Big). \tag{31}$$

The smoothness of the LQG cost $\mathscr{E}$ on the set $\mathbb{U}_0$ in (29) (which is an open subset of $\mathbb{U}$) suggests using gradient descent methods for the numerical solution of the CQLQG control problem (27).

## 7 Gradient Flow for the LQG Cost

The gradient descent approach to the solution of the CQLQG control problem is to move with the negative gradient flow for the LQG cost functional $\mathscr{E}$ in (27) towards a local minimum. The gradient descent can be regarded as a dynamical system governed by the ODE

$$\dot{u}(\tau) = -g(u(\tau)), \qquad u(0) = u_0. \tag{32}$$

Here, $\dot{(\,)} := \partial_\tau(\cdot)$ denotes the derivative with respect to fictitious time $\tau \geqslant 0$, and the gradient map $g : \mathbb{U}_0 \to \mathbb{U}$ is defined on the open set $\mathbb{U}_0$ from (29) as the Fréchet derivative

$$g(u) := \partial_u \mathscr{E}(u) = (\partial_R \mathscr{E}, \partial_b \mathscr{E}, \partial_e \mathscr{E}) \tag{33}$$

of the LQG cost $\mathscr{E}$ with respect to the matrix triple $u$ in the direct-sum Hilbert space $\mathbb{U}$ associated with the Hamiltonian parameterization of PR quantum controllers in (28). The initial point $u_0$ in (32) is assumed to satisfy

$$u_0 := (R_0, b_0, e_0) \in \mathbb{U}_0, \tag{34}$$

so that the corresponding PR controller is internally stabilizing. Unless $u_0$ is a stationary point of $\mathscr{E}$, the LQG cost is strictly decreasing along the trajectory of the ODE (32) in view of

$$\mathscr{E}(u)^{\bullet} = \langle g(u), \dot{u} \rangle = -\|g(u)\|^2. \tag{35}$$

The gradient of the LQG cost functional $\mathscr{E}$ is computed in the following lemma. To this end, we denote by $Q$ the observability Gramian of the pair $(\mathscr{A}, \mathscr{C})$ which is a unique solution of the ALE

$$\mathscr{A}^{\mathrm{T}} Q + Q\mathscr{A} + \mathscr{C}^{\mathrm{T}}\mathscr{C} = 0, \tag{36}$$

provided the matrix $\mathscr{A}$ in (11) is Hurwitz. Furthermore, use will be made of the Hankelian of the closed-loop system defined by

$$H := QP, \tag{37}$$

where $P$ is the controllability Gramian from (18). Also, $(2n \times 2n)$-matrices $X$ (such as $P$, $Q$, $H$) are partitioned into $(n \times n)$-blocks $X_{jk}$ as $X := \begin{bmatrix} X_{11} & X_{12} \\ X_{21} & X_{22} \end{bmatrix}$.

**Lemma 1** *For any $u \in \mathbb{U}_0$ from (29), the Fréchet derivatives of the LQG cost $\mathscr{E}$ in (27), which form the gradient $g$ in (33), can be computed as*

$$\partial_R \mathscr{E} = -2\mathrm{sym}(\Theta_2 H_{22}), \tag{38}$$
$$\partial_b \mathscr{E} = Q_{21} E d + Q_{22} b - \psi b J_2 - \chi d J_2, \tag{39}$$
$$\partial_e \mathscr{E} = H_{21} C^{\mathrm{T}} + Q_{21} B D^{\mathrm{T}} + Q_{22} e - \psi e D J_1 D^{\mathrm{T}}. \tag{40}$$

*Here, $\psi$ and $\chi$ are auxiliary $(n \times n)$-matrices defined by*

$$\psi := \mathrm{asym}(H_{22}\Theta_2^{-1}), \tag{41}$$
$$\chi := \Theta_2^{-1}(H_{12}^{\mathrm{T}} E + P_{21} F^{\mathrm{T}} G + P_{22} c^{\mathrm{T}} G^{\mathrm{T}} G), \tag{42}$$

*with P, Q, H the Gramians and the Hankelian from (18), (36), (37).* □

The proof of Lemma 1 is similar to that of [33, Theorem 1] and is given in Appendix A for completeness.

The following lemma employs the special structure of the LQG cost which prevents the trajectories of the gradient descent system (32) from "missing" the local minima once its stable equilibrium has been reached.

**Lemma 2** *A point $u_* \in \mathbb{U}_0$ in (29) is a stable equilibrium of the ODE (32) if and only if it is a local minimum of the LQG cost $\mathscr{E}$ in (27).* □

**PROOF.** The assertion of the lemma can be established by using [1, Theorem 3] and the *analyticity* [10] (rather than infinite differentiability) of the LQG cost $\mathscr{E}$ in a neighbourhood of any point $u \in \mathbb{U}_0$. The analyticity follows from the representation

$$\mathscr{E} = -\frac{1}{2}\mathrm{vec}(\mathscr{C}^{\mathrm{T}}\mathscr{C})^{\mathrm{T}}(\mathscr{A} \oplus \mathscr{A})^{-1}\mathrm{vec}(\mathscr{B}\mathscr{B}^{\mathrm{T}}) \tag{43}$$

which is obtained from (18), (27) by using the column-wise vectorization $\mathrm{vec}(\cdot)$ of matrices [17,29] and the Kronecker sum $\mathscr{A} \oplus \mathscr{A} := I_{2n} \otimes \mathscr{A} + \mathscr{A} \otimes I_{2n}$ of the matrix $\mathscr{A}$ with itself. Indeed, the representation (43) implies that $\mathscr{E}$ is a rational function of the entries of $\mathscr{A}$, $\mathscr{B}$, $\mathscr{C}$ in (11) which are, in turn, polynomial functions of the entries of $R$, $b$, $e$ in view of (24), (25), and hence, $\mathscr{E}(u)$ is a rational function of $u$. Therefore, the function $\mathscr{E}(u)$ is analytic on the open set $\mathbb{U}_0$ since the matrix $\mathscr{A} \oplus \mathscr{A}$ is also Hurwitz (and hence, nonsingular) for any $u \in \mathbb{U}_0$. ∎

The practical implementation of the gradient descent consists in solving the ODE (32) (whose right-hand side is provided by Lemma 1) by using a numerical algorithm equipped with a line search. Before proceeding to this approach in Section 10, we will discuss, in the next two sections, a symmetry property of the CQLQG control problem (27), which comes from the structure of linear quantum systems and can be taken into account in applications of gradient descent methods to such systems.

## 8 Symplectic Similarity Transformations and Equivalent Realizations of Coherent Quantum Controllers

The CQLQG control problem (27) inherits a special type of symmetry from the LQG cost $\mathscr{E}$ which is invariant under symplectic similarity transformations of the controller variables $\xi \mapsto \Sigma\xi$ and the corresponding transformations of the state-space matrices $a \mapsto \Sigma a \Sigma^{-1}$, $b \mapsto \Sigma b$, $e \mapsto \Sigma e$ and $c \mapsto c\Sigma^{-1}$ for any $\Sigma \in \mathbb{R}^{n\times n}$ satisfying

$$\Sigma\Theta_2\Sigma^{\mathrm{T}} = \Theta_2 \tag{44}$$

and thus preserving the CCR matrix $\Theta_2$ (see, for example, [31,33]). In what follows, it is assumed, without loss of generality, that the controller variables have a standard CCR matrix

$$\Theta_2 := I_{n/2} \otimes \mathbf{J}. \tag{45}$$

Therefore, the problem (27) is invariant with respect to the following linear transformations of the matrix triples $u \in \mathbb{U}$ of PR quantum controllers in (28):

$$u \mapsto (\Sigma^{-\mathrm{T}}R\Sigma^{-1}, \Sigma b, \Sigma e) =: \mathfrak{S}_\Sigma(u), \tag{46}$$

where $\Sigma$ is an arbitrary element of the Lie group of symplectic matrices $\mathrm{Sp}(n,\mathbb{R}) := \{\Sigma \in \mathbb{R}^{n\times n}:\ \Sigma\Theta_2\Sigma^{\mathrm{T}} = \Theta_2\}$ which is a connected non-compact subgroup of the special linear group $\mathrm{SL}(n,\mathbb{R}) := \{\Sigma \in \mathbb{R}^{n\times n}:\ \det\Sigma = 1\}$. We have used a slightly modified definition of the symplectic group which coincides with the standard definition up to the matrix transpose and corresponds to the CCR preservation (44). The nonlinear map $\Sigma \mapsto \mathfrak{S}_\Sigma$ in (46) is a group homomorphism since $\mathfrak{S}_{\Sigma_1} \circ \mathfrak{S}_{\Sigma_2} = \mathfrak{S}_{\Sigma_1\Sigma_2}$ for any symplectic matrices $\Sigma_1, \Sigma_2 \in \mathrm{Sp}(n,\mathbb{R})$. Accordingly, for any given $u \in \mathbb{U}$, we denote by

$$[u] := \{v \in \mathbb{U}:\ u \sim v\} \tag{47}$$

the class of equivalent representations of the coherent quantum controller parameterized by $u$. Here, the equivalence relation $u \sim v$ on the set $\mathbb{U}$ in (28) is understood in the sense of the transformation group (46) and means the existence of a matrix $\Sigma \in \mathrm{Sp}(n,\mathbb{R})$ such that

$$v = \mathfrak{S}_\Sigma(u).$$

Therefore, the equivalence class $[u]$ in (47) is the orbit of the Lie group $\mathfrak{S} := \big\{\mathfrak{S}_\Sigma(u):\ \Sigma \in \mathrm{Sp}(n,\mathbb{R})\big\}$ of symplectic transformations (46) acting on $u \in \mathbb{U}$ in (28):

$$[u] = \mathfrak{S}(u) := \big\{\mathfrak{S}_\Sigma(u):\ \Sigma \in \mathrm{Sp}(n,\mathbb{R})\big\}. \tag{48}$$

The set $[u]$ is a manifold in the Hilbert space $\mathbb{U}$. Its tangent space, which we denote by $\mathscr{T}(u)$, and the corresponding normal subspace $\mathscr{T}(u)^{\perp}$ (the orthogonal complement of $\mathscr{T}(u)$) are described by the following lemma.

**Lemma 3** *For any PR quantum controller specified by $u \in \mathbb{U}$ from (28), the tangent and normal subspaces to the equivalence*

*class* $[u]$ *in (48) can be represented as*

$$\mathscr{T}(u) = \big\{(-2\mathrm{sym}(R\Theta_2\varphi), \Theta_2\varphi b, \Theta_2\varphi e):\ \varphi \in \mathbb{S}_n\big\}, \qquad (49)$$

$$\mathscr{T}(u)^\perp = \big\{(\rho,\beta,\varepsilon) \in \mathbb{U}:\ 2R\rho - \beta b^{\mathrm{T}} - \varepsilon e^{\mathrm{T}} \in \Theta_2^{-1}\mathbb{A}_n\big\}. \qquad (50)$$

□

**PROOF.** For any fixed but otherwise arbitrary $u \in \mathbb{U}$, the first variation of the transformed matrix triple $\mathfrak{S}_\Sigma(u)$ in (46) with respect to the matrix $\Sigma$ about $I_n$ is computed as

$$\begin{aligned}\delta\mathfrak{S}_\Sigma(u) &= (\delta(\Sigma^{-\mathrm{T}})R\Sigma^{-1} + \Sigma^{-\mathrm{T}}R\delta(\Sigma^{-1}), (\delta\Sigma)b, (\delta\Sigma)e)\\ &= (-(\delta\Sigma)^{\mathrm{T}}R - R\delta\Sigma, (\delta\Sigma)b, (\delta\Sigma)e)\\ &= (-2\mathrm{sym}(R\delta\Sigma), (\delta\Sigma)b, (\delta\Sigma)e), \qquad \delta\Sigma \in \mathrm{sp}(n,\mathbb{R}).\end{aligned} \qquad (51)$$

Here,

$$\begin{aligned}\mathrm{sp}(n,\mathbb{R}) &:= \big\{\sigma \in \mathbb{R}^{n\times n}:\ \sigma\Theta_2 + \Theta_2\sigma^{\mathrm{T}} = 0\big\}\\ &= \big\{\Theta_2\varphi:\ \varphi \in \mathbb{S}_n\big\} = \Theta_2\mathbb{S}_n\end{aligned} \qquad (52)$$

is the subspace of Hamiltonian matrices which form the Lie algebra [22] of infinitesimal generators for the symplectic group $\mathrm{Sp}(n,\mathbb{R})$. In (51), use is also made of the relation $\delta(\Sigma^{-1}) = -\Sigma^{-1}(\delta\Sigma)\Sigma^{-1} = -\delta\Sigma$ (evaluated at $\Sigma = I_n$), and the fact that $(\delta\Sigma)\Sigma^{-1} \in \mathrm{sp}(n,\mathbb{R})$ for symplectic matrices $\Sigma \in \mathrm{Sp}(n,\mathbb{R})$. Furthermore, the identity $\sigma^{\mathrm{T}}R + R\sigma = 2\mathrm{sym}(R\sigma)$ follows from the symmetry of the matrix $R$. A combination of (51) with (52) leads to (49), where the matrices $\varphi \in \mathbb{S}_n$ parameterize the infinitesimal generators of the symplectic similarity transformation group $\mathfrak{S}$ acting on $u$. In order to prove (50), we note that a matrix triple $w := (\rho,\beta,\varepsilon) \in \mathbb{U}$ is orthogonal to the tangent space $\mathscr{T}(u)$ in (49) if and only if the corresponding inner product (30) vanishes for any $\varphi \in \mathbb{S}_n$:

$$\begin{aligned}\langle(-2\mathrm{sym}(R\Theta_2\varphi), \Theta_2\varphi b, \Theta_2\varphi e), w\rangle &= \langle -2\mathrm{sym}(R\Theta_2\varphi), \rho\rangle + \langle\Theta_2\varphi b, \beta\rangle + \langle\Theta_2\varphi e, \varepsilon\rangle\\ &= \langle\varphi, 2\mathrm{sym}(\Theta_2 R\rho)\rangle - \langle\varphi, \mathrm{sym}(\Theta_2\beta b^{\mathrm{T}})\rangle - \langle\varphi, \mathrm{sym}(\Theta_2\varepsilon e^{\mathrm{T}})\rangle\\ &= \langle\varphi, \mathrm{sym}(\Theta_2(2R\rho - \beta b^{\mathrm{T}} - \varepsilon e^{\mathrm{T}}))\rangle = 0,\end{aligned} \qquad (53)$$

where use is made of the antisymmetry of the CCR matrix $\Theta_2$. The fulfillment of (53) for all $\varphi \in \mathbb{S}_n$ is equivalent to the relation $\mathrm{sym}(\Theta_2(2R\rho - \beta b^{\mathrm{T}} - \varepsilon e^{\mathrm{T}})) = 0$, and hence, $w \in \mathscr{T}(u)^\perp$ if and only if the matrix $2R\rho - \beta b^{\mathrm{T}} - \varepsilon e^{\mathrm{T}}$ is skew Hamiltonian in the sense of the inclusion in (50). ■

Associated with every $u \in \mathbb{U}$ is the orthogonal decomposition of the Hilbert space $\mathbb{U}$ into the subspaces $\mathscr{T}(u)$ and $\mathscr{T}(u)^\perp$ in (49) and (50):

$$\mathbb{U} = \mathscr{T}(u) \oplus \mathscr{T}(u)^\perp. \qquad (54)$$

Note that the tangent space $\mathscr{T}(u)$ is a proper subspace of $\mathbb{U}$ in view of the relations $\dim\mathscr{T}(u) \leqslant \dim\mathbb{S}_n < \dim\mathbb{U}$ which are obtained by combining (49) with (31). Therefore, the normal subspace $\mathscr{T}(u)^\perp$ is nontrivial, and its dimension admits the lower bound

$$\dim\mathscr{T}(u)^\perp = \dim\mathbb{U} - \dim\mathscr{T}(u) \geqslant n(m_2 + p_1).$$

Now, for any stabilizing PR quantum controller specified by a matrix triple $u \in \mathbb{U}_0$ from (29), the LQG cost $\mathscr{E}$ in (27) is invariant under the symplectic similarity transformations (46) in the sense that

$$\mathscr{E}(\mathfrak{S}_\Sigma(u)) = \mathscr{E}(u) \qquad (55)$$

for any symplectic matrix $\Sigma \in \mathrm{Sp}(n,\mathbb{R})$. That is, $\mathscr{E}$ is constant on every equivalence class $[u]$ given by (48). In accordance with general results on quotient manifolds (see, for example, [2, p. 49]), the symplectic invariance of the LQG cost leads to the following "orientation" of its gradient in terms of the decomposition (54).

**Lemma 4** *For any* $u \in \mathbb{U}_0$*, the gradient* $g(u)$ *of the LQG cost* $\mathscr{E}$ *in (33) is orthogonal to the tangent space (49) generated by* *the group of symplectic similarity transformations (46):*

$$g(u) \in \mathscr{T}(u)^{\perp}. \tag{56}$$

*Equivalently, the partial Fréchet derivatives $\partial_R \mathscr{E}$, $\partial_b \mathscr{E}$, $\partial_e \mathscr{E}$ satisfy*

$$2R\partial_R \mathscr{E} - \partial_b \mathscr{E} b^{\mathrm{T}} - \partial_e \mathscr{E} e^{\mathrm{T}} \in \Theta_2^{-1}\mathbb{A}_n. \tag{57}$$

□

**PROOF.** Consider a symplectic matrix $\Sigma$, which is given by

$$\Sigma := \mathrm{e}^{\lambda\sigma}, \qquad \sigma := \Theta_2 \varphi \tag{58}$$

and depends on a scalar parameter $\lambda \in \mathbb{R}$. Here, $\sigma$ is a Hamiltonian matrix specified by a fixed but otherwise arbitrary matrix $\varphi \in \mathbb{S}_n$. Since we are concerned with a matrix group, the exponential map in (58) is the usual matrix exponential whose derivative $(\cdot)' := \partial_\lambda(\cdot)|_{\lambda=0}$ is computed as

$$\Sigma' = \sigma. \tag{59}$$

For any given $u \in \mathbb{U}_0$, the left-hand side of the relation (55) is a composite function $\lambda \mapsto \Sigma \mapsto \mathfrak{S}_\Sigma(u) \mapsto \mathscr{E}(\mathfrak{S}_\Sigma(u))$, while the right-hand side of (55) is a constant. Hence, by differentiating both sides of (55) with respect to $\lambda$ and using the chain rule, it follows that

$$\begin{aligned} 0 &= \big\langle g(u),\, \mathfrak{S}_\Sigma(u)' \big\rangle \\ &= \big\langle (\partial_R \mathscr{E}, \partial_b \mathscr{E}, \partial_e \mathscr{E}),\, (-2\mathrm{sym}(R\sigma), \sigma b, \sigma e) \big\rangle \\ &= \big\langle \mathrm{sym}(\Theta_2(2R\partial_R \mathscr{E} - \partial_b \mathscr{E} b^{\mathrm{T}} - \partial_e \mathscr{E} e^{\mathrm{T}})),\, \varphi \big\rangle, \end{aligned} \tag{60}$$

where the third equality is obtained similarly to (53). Here, in view of (58), (59) and in accordance with (51), the directional derivative

$$\mathfrak{S}_\Sigma(u)' = (-2\mathrm{sym}(R\sigma), \sigma b, \sigma e) \tag{61}$$

belongs to the tangent space $\mathscr{T}(u)$ in (49). The orthogonality (56) now follows from the first equality in (60) and the fact that the directional derivatives $\mathfrak{S}_\Sigma(u)'$ in (61), considered for all possible matrices $\varphi \in \mathbb{S}_n$, exhaust the subspace $\mathscr{T}(u)$. The relation (56) implies (in fact, is equivalent to) (57) in view of (50). Alternatively, (57) can also be obtained by noting that the fulfillment of (60) for any $\varphi \in \mathbb{S}_n$ leads to

$$\mathrm{sym}(\Theta_2(2R\partial_R \mathscr{E} - \partial_b \mathscr{E} b^{\mathrm{T}} - \partial_e \mathscr{E} e^{\mathrm{T}})) = 0, \tag{62}$$

which is equivalent to the matrix $2R\partial_R \mathscr{E} - \partial_b \mathscr{E} b^{\mathrm{T}} - \partial_e \mathscr{E} e^{\mathrm{T}}$ being skew Hamiltonian in the sense of (57). ■

Note that the relation (57) (whose equivalent representation is given by (62)) originates from the symplectic invariance (55) of the LQG cost $\mathscr{E}$ regardless of its particular form. An equivalent interpretation of (56) is that the tangent space $\mathscr{T}(u)$ is contained by the orthogonal hyperplane $g(u)^{\perp}$ (which is the tangent hyperplane for the level set $\mathscr{E} = \mathrm{const}$, provided $g(u) \neq 0$).

## 9 Norm-balanced Realizations of Coherent Quantum Controllers

Except for trivial cases (such as $u = 0$), the equivalence class $[u]$ in (47) is, in general, an unbounded subset of $\mathbb{U}$ and may contain state-space realizations with large matrices which are difficult to handle numerically. In particular, such representations of a coherent quantum controller may lead to a loss of accuracy in computing the LQG cost and its derivatives. This suggests using the squared norm on the Hilbert space $\mathbb{U}$ in order to penalize "pathological" representatives of the equivalence class $[u]$. To this end, a matrix triple $u \in \mathbb{U}$ in (28) is said to be a *norm-balanced realization* of the corresponding coherent quantum controller, if $u$ is a stationary point of the auxiliary minimization problem

$$\mathscr{N}(v) := \frac{1}{2}\|v\|^2 \longrightarrow \min, \qquad v \in [u]. \tag{63}$$

Here, the factor $\frac{1}{2}$ is introduced for further convenience, and $\|\cdot\|$ is the norm associated with the direct-sum inner product in (30).

**Lemma 5** *If $u \in \mathbb{U}_0$ is a norm-balanced realization of a stabilizing PR quantum controller, then this property is preserved by the gradient flow (32).* □

**PROOF.** The stationarity condition, associated with (63), is equivalent to the gradient of $\mathcal{N}(u)$ being orthogonal to the tangent space $\mathcal{T}(u)$ in (49):

$$\partial_u \mathcal{N}(u) = u \in \mathcal{T}(u)^{\perp}. \tag{64}$$

By applying (50) of Lemma 3 to the matrices $\rho = R$, $\beta = b$ and $\varepsilon = e$, it follows that the inclusion (64) is satisfied if and only if $2R^2 - bb^{\mathrm{T}} - ee^{\mathrm{T}} \in \Theta_2^{-1}\mathbb{A}_n$ or, equivalently,

$$\mathrm{sym}(\Theta_2(2R^2 - bb^{\mathrm{T}} - ee^{\mathrm{T}})) = 0. \tag{65}$$

If the matrix triple $u \in \mathbb{U}_0$ (not necessarily norm-balanced) is evolved by the gradient flow (32), then the corresponding derivative of the left-hand side of (65) with respect to the fictitious time parameter $\tau$ is

$$\begin{aligned}\mathrm{sym}(\Theta_2(2R^2 - bb^{\mathrm{T}} - ee^{\mathrm{T}}))^{\bullet} &= \mathrm{sym}(\Theta_2\Xi) + \mathrm{sym}(\Theta_2\Xi^{\mathrm{T}})\\ &= \mathrm{sym}(\Theta_2\Xi) + \Theta_2\mathrm{sym}(\Theta_2\Xi)\Theta_2 = 0,\end{aligned} \tag{66}$$

where

$$\Xi := 2R\dot{R} - \dot{b}b^{\mathrm{T}} - \dot{e}e^{\mathrm{T}} = -(2R\partial_R\mathcal{E} - \partial_b\mathcal{E}b^{\mathrm{T}} - \partial_e\mathcal{E}e^{\mathrm{T}})$$

is a skew Hamiltonian matrix in view of (57). Here, use is also made of the identity $\mathrm{sym}(\Theta_2\Xi^{\mathrm{T}}) = \Theta_2\mathrm{sym}(\Theta_2\Xi)\Theta_2$ which holds for an arbitrary matrix $\Xi \in \mathbb{R}^{n\times n}$ since the CCR matrix $\Theta_2$ in (45) satisfies $\Theta_2^2 = -I_n$. Now, if $u \in \mathbb{U}_0$ is a given norm-balanced realization of a stabilizing coherent quantum controller, then the fulfillment of (65) propagates along the gradient flow in view of (66). ■

The proof of Lemma 5 shows that the left-hand side of (65) is a first integral of motion (a conserved quantity) for the gradient flow (32) irrespective of $u$ being a norm-balanced realization. Since this property is a corollary of the symplectic invariance (rather than a particular form) of the LQG cost, the conserved quantity $\mathrm{sym}(\Theta_2(2R^2 - bb^{\mathrm{T}} - ee^{\mathrm{T}}))$ is similar to the Noether invariants [22].

In view of (46), (48) and (65), finding a norm-balanced realization in the equivalence class $[u]$ reduces to finding a symplectic matrix $\Sigma \in \mathrm{Sp}(n,\mathbb{R})$ such that

$$\mathrm{sym}(\Theta_2(2(\Sigma^{-\mathrm{T}}R\Sigma^{-1})^2 - \Sigma(bb^{\mathrm{T}} + ee^{\mathrm{T}})\Sigma^{\mathrm{T}})) = 0.$$

Instead of solving this nonlinear algebraic equation, an alternative way to employ the idea of norm balancing (63) is to modify the gradient flow (32) as

$$\dot{u} = \gamma(u) - g(u), \tag{67}$$

where $\gamma(u) \in \mathcal{T}(u)$ takes values in the tangent space (49). Due to the orthogonality (56), the LQG cost $\mathcal{E}$ decreases at the same rate (35) along trajectories of the modified flow (67):

$$\mathcal{E}(u)^{\bullet} = \langle g(u), \gamma(u) - g(u)\rangle = -\|g(u)\|^2.$$

However, (67) provides flexibility in controlling the value of $\mathcal{N}(u)$ in (63) during the gradient descent. In particular, $\gamma(u)$ can be chosen so as to keep it constant, which is equivalent to

$$\mathcal{N}(u)^{\bullet} = \langle u, \gamma(u) - g(u)\rangle = 0. \tag{68}$$

A solution $\gamma(u) \in \mathcal{T}(u)$ of the equation (68), which minimizes the norm $\|\gamma(u)\|$, can be computed in terms of the decomposition (54) as

$$\gamma(u) = \frac{1}{\|\Pi_u(u)\|^2}\langle \Pi_u^{\perp}(u), g(u)\rangle \Pi_u(u). \tag{69}$$

Here, $\Pi_u : \mathbb{U} \to \mathscr{T}(u)$ denotes the operator of orthogonal projection onto the tangent space $\mathscr{T}(u)$. Accordingly, the operator

$$\Pi_u^\perp := \mathscr{I} - \Pi_u \tag{70}$$

describes the orthogonal projection onto the normal subspace $\mathscr{T}(u)^\perp$, where $\mathscr{I}$ is the identity operator on $\mathbb{U}$. The denominator in (69) vanishes if and only if the inclusion (64) holds, that is, if $u$ is a norm-balanced realization of the coherent quantum controller. Otherwise, (69) is well defined, and (67) takes the form

$$\dot{u} = -\Big(\mathscr{I} - \frac{1}{\|\Pi_u(u)\|^2}\Pi_u(u)\langle \Pi_u^\perp(u), \cdot\rangle\Big)(g(u)). \tag{71}$$

The fact that the right-hand side of (71) is organized as a linear operator acting on the gradient $g(u)$ makes the modified gradient flow reminiscent of the steepest descent with respect to a different Riemannian metric [2], except that this operator is not self-adjoint.

## 10 A Gradient Descent Algorithm with an Adaptive Line Search

We will now consider a numerical algorithm which implements the gradient descent method (32) for the CQLQG control problem (27) in the form

$$u_{k+1} := u_k - s_k g(u_k), \qquad k = 0, 1, 2, \ldots. \tag{72}$$

This recurrence equation is initialized with matrices $R_0$, $b_0$, $e_0$ of an internally stabilizing PR controller in (34) (see Section 10.1). The gradient $g(u_k)$ is computed by using Lemma 1, and the stepsize $s_k > 0$ is chosen as described in Section 10.2. The iterations in (72) are stopped when a termination condition is satisfied (see Section 10.3). The ingredients of the algorithm are discussed in the subsequent sections.

### *10.1 Initialization*

The initialization of the gradient descent algorithm (72) requires an internally stabilizing PR quantum controller. The existence of such controllers (that is, nonemptiness of the set $\mathbb{U}_0$ in (29)) for a given quantum plant (and a systematic method of finding them) remains an open problem. In the present version of the algorithm, the quantum stabilization problem is solved by using a pure random search in the finite-dimensional Hilbert space $\mathbb{U}$ in (28) (that is, the triples $(R, b, e)$ are randomly generated until the corresponding closed-loop matrix $\mathscr{A}$ becomes Hurwitz).

### *10.2 Stepsize selection*

According to the conventional limited minimization rule (see, for example, [4]), the stepsize $s_k$ is chosen for each iteration of the gradient descent by solving the minimization problem

$$s_k \in \operatorname*{Arg\,min}_{0 \leqslant s \leqslant h_k} \mathscr{E}(u_k - s g(u_k)) \tag{73}$$

with a *constant* search horizon $h_k := h > 0$. Here, we use the convention that $\mathscr{E}(u) := +\infty$ if $u \not= \mathbb{U}_0$ (thus discarding those controllers which are not internally stabilizing). A restricted version of the line-search with a constant horizon $h$ may suffer from the inability to adapt properly to the behaviour of the function $\mathscr{E}$ in its minimization over the "ray" $\{u_k - s g(u_k) :\ s \geqslant 0\} \bigcap \mathbb{U}_0$. In order to overcome this issue, for the stepsize selection in the gradient descent algorithm (72), we will use a modified version of the limited minimization rule with an adaptive choice of the search horizon $h_k$ in each iteration. More precisely, $h_k$ can be chosen so as to enable (73) to "capture" the minimum of $\mathscr{E}$ over the whole ray if $\mathscr{E}$ were a strictly convex quadratic function. To this end, consider the following quadratic truncation of the Taylor series

$$\mathscr{E}(u - sg) = \mathscr{E}(u) - s\mathscr{D}_g\mathscr{E} + \frac{s^2}{2}\mathscr{D}_g^2\mathscr{E} + o(s^2), \tag{74}$$

where

$$\mathscr{D}_v\mathscr{E}(u) := \partial_s\mathscr{E}(u + sv)\big|_{s=0} = \langle g(u), v\rangle, \tag{75}$$

$$\mathscr{D}_v^2\mathscr{E}(u) := \partial_s^2\mathscr{E}(u + sv)\big|_{s=0} = \langle \partial_u^2\mathscr{E}(u)(v), v\rangle, \tag{76}$$

are the first and second-order Gâteaux (or directional) derivatives [13] of the LQG cost at a point $u \in \mathbb{U}_0$ (specifying an internally stabilizing controller) along $v \in \mathbb{U}$. The second-order Gâteaux derivative $\mathscr{D}_v^2 \mathscr{E}(u)$ in (76) is computed in Appendix B and employs the second-order Fréchet derivative (the Hessian operator) $\partial_u^2 \mathscr{E}(u) := \partial_u g(u)$ which is a self-adjoint operator on the Hilbert space $\mathbb{U}$ in (28). In application to the gradient direction $v = g(u)$, the Gâteaux derivative (75) takes the form

$$\mathscr{D}_g \mathscr{E} = \|g\|^2 = \|\partial_R \mathscr{E}\|^2 + \|\partial_b \mathscr{E}\|^2 + \|\partial_e \mathscr{E}\|^2 \geqslant 0, \tag{77}$$

which was already used in (35). Now, if $\mathscr{D}_g^2 \mathscr{E}(u) > 0$, then the quadratic polynomial of $s$ on the right-hand side of (74) (with the higher-order terms being neglected) achieves its unique minimum at a nonnegative value of $s$:

$$\arg\min_{s \geqslant 0} \left( \frac{s^2}{2} \mathscr{D}_g^2 \mathscr{E} - s \mathscr{D}_g \mathscr{E} \right) = \frac{\mathscr{D}_g \mathscr{E}}{\mathscr{D}_g^2 \mathscr{E}} = \frac{\|g\|^2}{\mathscr{D}_g^2 \mathscr{E}}. \tag{78}$$

This suggests using the right-hand side of (78) as a search horizon $h_k$ in (73), provided $\mathscr{D}_g^2 \mathscr{E}(u) > 0$. However, if the latter inequality does not hold, the argument, based on a quadratic approximation of the minimization problem (73), is no longer valid and needs to be amended. In this case (when $\mathscr{D}_g^2 \mathscr{E}(u) \leqslant 0$), the search horizon can be chosen so as to avoid the domination of nonlinear terms over the linear term in the quadratically truncated Taylor series for the LQG cost along the ideal gradient descent trajectory in (32):

$$\begin{aligned} \mathscr{E}(u(\tau + s)) &= \mathscr{E}(u(\tau)) + \mathscr{E}(u)^{\bullet} s + \mathscr{E}(u)^{\bullet\bullet} \frac{s^2}{2} + o(s^2) \\ &= \mathscr{E}(u(\tau)) - \|g\|^2 s + \mathscr{D}_g^2 \mathscr{E} s^2 + o(s^2). \end{aligned} \tag{79}$$

Here, the derivatives

$$\begin{aligned} \mathscr{E}(u)^{\bullet} &= -\mathscr{D}_g \mathscr{E} = -\|g\|^2, \\ \mathscr{E}(u)^{\bullet\bullet} &= -2\langle g, \dot{g} \rangle = 2\langle g, \partial_u^2 \mathscr{E}(u)(g) \rangle = 2\mathscr{D}_g^2 \mathscr{E} \end{aligned}$$

are evaluated by regarding $\mathscr{E}(u(\tau))$ and $g(u(\tau))$ as composite functions of the fictitious time parameter $\tau$ with the aid of (75)–(77). For $s \geqslant 0$, the comparison of the absolute values $\|g\|^2 s$ and $|\mathscr{D}_g^2 \mathscr{E}| s^2$ of the linear and quadratic terms in (79) shows that the latter does not dominate the former if

$$s \leqslant \frac{\|g\|^2}{|\mathscr{D}_g^2 \mathscr{E}|}. \tag{80}$$

This inequality is closely related to the accuracy of (72) as an Euler scheme for the numerical integration of the ODE (32). More precisely, if the stepsizes $s_k > 0$ in (72) are significantly smaller than the respective values of the right-hand side of (80), then $u_k$ becomes an accurate approximation of the ideal gradient descent trajectory $u(\tau)$ at fictitious time $\tau := s_0 + \ldots + s_{k-1}$. A combination of (78) and (80) justifies the following heuristic rule for choosing the search horizon at the current point $u_k \in \mathbb{U}_0$:

$$h_k := \min \left( h_{\max}, \frac{\|g\|^2}{|\mathscr{D}_g^2 \mathscr{E}|} \Big|_{u = u_k} \right). \tag{81}$$

Here, $h_{\max}$ is a given positive threshold which becomes active, for example, if $\mathscr{D}_g^2 \mathscr{E}$ vanishes. The stepsize selection algorithm, considered below, replaces the minimization problem in (73) with a different procedure which involves a finite subset of values of $s$ from a geometric progression

$$s_{k,\mu} := h_k f^{\mu}, \qquad \mu = 0, 1, 2, \ldots \tag{82}$$

whose initial value $h_k$ is given by (81). The common ratio $f \in (0,1)$ is a parameter of the algorithm which affects how "densely" the progression fills the interval $[0, h_k]$. Now, the adaptive stepsize selection algorithm proceeds as follows:

$$s_k := s_{k,j}, \tag{83}$$

where the $j$th element of the geometric progression (82) is chosen according to the Armijo rule [4] with a parameter $\sigma \in (0,1)$:

$$j := \min \{ \mu \geqslant 0 : \mathscr{E}(u_k) - \mathscr{E}(u_k - s_{k,\mu} g(u_k)) \geqslant \sigma s_{k,\mu} \|g(u_k)\|^2 \}. \tag{84}$$

Here, the subset of indices $\mu$ is nonempty in view of (79) since $\sigma < 1$ and $\lim_{\mu\to+\infty} s_{k,\mu} = 0$. The inequality in (84) is important in the convergence analysis of the gradient descent algorithm. In particular, the condition $\sigma > 0$ ensures that $\mathscr{E}(u_k)$ is strictly decreasing until $u_k$ achieves a stationary point of the LQG cost. Such a point is a stable equilibrium of the gradient descent only if it delivers a local minimum to the LQG cost (cf. Lemma 2).

### *10.3 Termination condition*

Since the gradient descent sequence $u_k$ in (72) can converge to a local minimum of the LQG cost only asymptotically, as $k \to +\infty$, the algorithm is equipped with a termination condition (for stopping the iterations) which reflects the proximity to the limit point. More precisely, we use the following termination condition which employs the relative smallness of the gradient as specified by a dimensionless parameter $\varepsilon > 0$:

$$s_k \|g(u_k)\| \leqslant \varepsilon \|u_k\|. \tag{85}$$

## 11 Convergence of the Gradient Descent Algorithm

Since the proposed algorithm is based on the classical gradient descent approach, its convergence analysis follows a similar reasoning which is provided below for completeness.

**Theorem 6** *Suppose $(u_k)_{k\geqslant 0}$ is the gradient descent sequence in (72) with the stepsize selection described by (81)–(84). Then every limit point $u_* \in \mathbb{U}_0$ of this sequence is a stationary point of the LQG cost $\mathscr{E}$, that is, $g(u_*) = 0$.* □

**PROOF.** Since the sequence $\mathscr{E}(u_k) \geqslant 0$ is nonincreasing, it has a finite limit. Therefore, (72), (83) and the Armijo rule (84) imply that $\sigma s_k \|g(u_k)\|^2 \leqslant \mathscr{E}(u_k) - \mathscr{E}(u_{k+1}) \to 0$ as $k \to +\infty$. Hence, in view of $\sigma > 0$, it follows that

$$\lim_{k\to+\infty} \big(s_k \|g(u_k)\|^2\big) = 0. \tag{86}$$

Now, assume that the gradient descent sequence $u_k$ has a limit point $u_* := \lim_{\mathscr{K}\ni k\to+\infty} u_k \in \mathbb{U}_0$ such that $g(u_*) \neq 0$, where $\mathscr{K} := \{0 \leqslant k_1 < k_2 < \ldots\}$ is an infinite subset of nonnegative integers which specifies the corresponding subsequence of $u_k$. Then the analyticity of the LQG cost on the open set $\mathbb{U}_0$ implies that

$$\lim_{\mathscr{K}\ni k\to+\infty} g(u_k) = g(u_*) \neq 0, \tag{87}$$

$$\lim_{\mathscr{K}\ni k\to+\infty} h_k = \min\left(h_{\max}, \frac{\|g(u_*)\|^2}{|\mathscr{D}_g^2 \mathscr{E}(u_*)|}\right) > 0, \tag{88}$$

where use is made of (81). Note that, if $\mathscr{D}_g^2 \mathscr{E}(u_*) = 0$, then the limit in (88) is equal to $h_{\max} > 0$. A combination of (87) with (86) implies that

$$\lim_{\mathscr{K}\ni k\to+\infty} s_k = 0. \tag{89}$$

In turn, by combining (89) with (88) and recalling (83) together with the condition $0 < f < 1$, it follows that the indices $j_p := \log_f \frac{s_{k_p}}{h_{k_p}} = \frac{\ln h_{k_p} - \ln s_{k_p}}{-\ln f}$ of the elements of the geometric progression in (82), which correspond to $k_p \in \mathscr{K}$, diverge to infinity as $p \to +\infty$, and hence, $j_p \geqslant 1$ for all sufficiently large $p$. For all such $p$, the stepsize candidates $s_{k_p, j_p - 1} = \frac{s_{k_p}}{f}$ do not pass the Armijo selection rule (84), that is,

$$\mathscr{E}(u_k) - \mathscr{E}\Big(u_k - \frac{s_k}{f} g(u_k)\Big) < \sigma \frac{s_k}{f} \|g(u_k)\|^2 \tag{90}$$

for all sufficiently large $k \in \mathscr{K}$. Upon multiplying both parts of (90) by $\frac{f}{s_k}$ and taking the limit, this inequality leads to

$$\begin{aligned}\|g(u_*)\|^2 &= \lim_{\mathscr{K}\ni k\to+\infty} \Big(\frac{f}{s_k}\Big(\mathscr{E}(u_k) - \mathscr{E}\big(u_k - \frac{s_k}{f} g(u_k)\big)\Big)\Big)\\ &\leqslant \sigma \lim_{\mathscr{K}\ni k\to+\infty} \|g(u_k)\|^2 = \sigma \|g(u_*)\|^2,\end{aligned} \tag{91}$$

where use is also made of (87) and (89). However, since $\sigma < 1$, the inequality in (91) contradicts the assumption that $g(u_*) \neq 0$. This contradiction shows that any limit point $u_* \in \mathbb{U}_0$ of the gradient descent sequence satisfies $g(u_*) = 0$. ■

In view of the symplectic invariance of the LQG cost, discussed in Section 8, its stationary points (including local minima of the CQLQG control problem (27)) are not isolated. In fact, if $u_* \in \mathbb{U}_0$ is a stationary point (in particular, a local minimum) of the LQG cost $\mathscr{E}$, then so also is every representative of the equivalence class $[u_*]$ in (48). Furthermore, if $u_*$ is a local minimum of the LQG cost, then the Hessian operator $\partial_u^2 \mathscr{E}(u_*)$ is positive semi-definite and its null space contains the corresponding tangent space from (49):

$$\mathscr{T}_* := \mathscr{T}(u_*) \subset \ker \partial_u^2 \mathscr{E}(u_*). \tag{92}$$

The fact that $\partial_u^2 \mathscr{E}(u_*)$ is not positive definite on the space $\mathbb{U}$ complicates the asymptotic convergence rate analysis for the gradient descent algorithm of Section 10 (see, for example, [30] and references therein). If the inclusion (92) is an equality, that is,

$$\mathscr{T}_* = \ker \partial_u^2 \mathscr{E}(u_*), \tag{93}$$

then, in view of the decomposition (54), the normal subspace $\mathscr{T}_*^\perp$ is invariant with respect to the Hessian operator and, moreover, $\partial_u^2 \mathscr{E}(u_*)$ is positive definite on $\mathscr{T}_*^\perp$ and hence, $\partial_u^2 \mathscr{E}(u_*)(\mathscr{T}_*^\perp) = \mathscr{T}_*^\perp$. The positive definiteness of $\partial_u^2 \mathscr{E}(u_*)$ on the subspace $\mathscr{T}_*^\perp$ implies that the corresponding second-order Gâteaux derivative in (76) is strictly positive on the punctured normal subspace: $\mathscr{D}_v^2 \mathscr{E}(u_*) > 0$ for all $v \in \mathscr{T}_*^\perp \setminus \{0\}$. This suggests the existence of a sufficiently small neighbourhood $\mho \subset \mathbb{U}_0$ of the local minimum $u_*$ and a constant $\ell > 0$ such that the inequalities

$$\frac{\ell}{2} \vartheta(u)^2 \leqslant \mathscr{E}(u) - \mathscr{E}(u_*) \leqslant \frac{1}{2\ell} \|g(u)\|^2, \tag{94}$$

are satisfied for all $u \in \mho$, where

$$\vartheta(u) := \inf_{v \in [u_*]} \|u - v\| \tag{95}$$

denotes the deviation of $u$ from the equivalence class $[u_*]$ (of locally optimal controllers) associated with $u_*$ by (47). Then, assuming, without loss of generality, that the set $\mho$ is convex, it follows that

$$\mathscr{E}(u) - \mathscr{E}(u_*) \leqslant \frac{L}{2} \|u - u_*\|^2 \tag{96}$$

holds for all $u \in \mho$ with another positive constant $L$. The role of such $L$ can be played by an upper bound for the largest eigenvalue of the Hessian operator:

$$L := \sup_{u \in \mho} \lambda_{\max}(\partial_u^2 \mathscr{E}(u)). \tag{97}$$

For completeness of exposition, we note that (97) indeed guarantees the fulfillment of (96) due to convexity of $\mho$ and the integral representation

$$\begin{aligned}
\mathscr{E}(u) - \mathscr{E}(u_*) &= \int_0^1 \langle g(\widehat{u}_\lambda), u - u_* \rangle \mathrm{d}\lambda \\
&= \int_{[0,1]^2} \lambda \langle \partial_u^2 \mathscr{E}(\widehat{u}_{\lambda\mu})(u - u_*), u - u_* \rangle \mathrm{d}\lambda \mathrm{d}\mu \\
&= \int_{[0,1]^2} \lambda \mathscr{D}_{u-u_*}^2 \mathscr{E}(\widehat{u}_{\lambda\mu}) \mathrm{d}\lambda \mathrm{d}\mu \\
&\leqslant L \int_{[0,1]^2} \lambda \mathrm{d}\lambda \mathrm{d}\mu \|u - u_*\|^2 = \frac{L}{2} \|u - u_*\|^2.
\end{aligned}$$

Here, use is made of (76), and $\widehat{u}_\lambda := (1 - \lambda) u_* + \lambda u \in \mho$ is an intermediate point of the line segment with the endpoints $u_*$ and $u$, which is contained by $\mho$. A comparison of (96) with the first inequality in (94) shows that $\ell \leqslant L$. Furthermore, the truncated Taylor series expansions

$$\begin{aligned}
g(u) &= \partial_u^2 \mathscr{E}(u_*)(\Pi_*^\perp(u - u_*)) + o(\|u - u_*\|), \\
\mathscr{E}(u) &= \mathscr{E}(u_*) + \frac{1}{2} \mathscr{D}_{\Pi_*^\perp(u-u_*)}^2 \mathscr{E}(u_*) + o(\|u - u_*\|^2), \qquad \text{as } u \to u_*,
\end{aligned}$$

suggest that, in the case (93), the constants $\ell$ and $L$ in (94) and (97) can be chosen close to the smallest and largest eigenvalues of the Hessian operator $\partial_u^2 \mathscr{E}(u_*)$ on the corresponding normal subspace $\mathscr{T}_*^\perp$ (provided the neighbourhood $\mho$ is small enough). Here,

$$\Pi_* := \Pi_{u_*}$$

denotes the orthogonal projection onto the tangent space $\mathscr{T}_*$ at the local minimum $u_*$, so that $\Pi_*^\perp = \mathscr{I} - \Pi_*$ is the orthogonal projection onto $\mathscr{T}_*^\perp$ in accordance with (70).

**Theorem 7** *Suppose the gradient descent sequence $(u_k)_{k\geqslant 0}$ in (72), with the gradient $g$ of the LQG cost $\mathscr{E}$ computed in Lemma 1 and the stepsize selection following the Armijo rule (81)–(84), converges to a local minimum $u_* \in \mathbb{U}_0$ of the CQLQG control problem (27). Also, assume that the second-order Fréchet derivative $\partial_u^2 \mathscr{E}$ satisfies (93) and that (94) is fulfilled in a small convex neighbourhood $\mho$ of $u_*$ with a positive constant $\ell \leqslant L$, where $L$ is given by (97). Then*

$$\mathscr{E}(u_k) - \mathscr{E}(u_*) = O(r^k), \qquad \vartheta(u_k) = O(r^{k/2}), \qquad \text{as } k \to +\infty, \tag{98}$$

*where use is made of (95), and $0 < r < 1$ is computed in terms of the parameters $f$ and $\sigma$ of the Armijo rule and the constants $\ell$ and $L$ as*

$$r := 1 - 4f\sigma(1-\sigma)\frac{\ell}{L}. \tag{99}$$

□

**PROOF.** Since the sequence $(u_k)_{k\geqslant 0}$ converges to $u_*$, then, in view of (81), both $u_k$ and $u_k - h_k g(u_k)$, for all sufficiently large $k$, are in a sufficiently small convex neighbourhood $\mho$ of $u_*$ for which (94) and (96) are satisfied. For such $k$, the second-order expansion of the LQG cost $\mathscr{E}$ leads to

$$\mathscr{E}(u_k) - \mathscr{E}(u_k - s_{k,\mu} g(u_k)) = s_{k,\mu} \|g(u_k)\|^2 - \frac{1}{2} s_{k,\mu}^2 \mathscr{D}_{g(u_k)}^2 \mathscr{E}(\widehat{u}_{k,\mu}). \tag{100}$$

Here, $\widehat{u}_{k,\mu}$ is an intermediate point of the line segment joining $u_k$ and $u_k - s_{k,\mu} g(u_k)$, and use is made of $s_{k,\mu}$ from (82) with $0 < f < 1$. This line segment is contained by $\mho$. Hence, by combining (100) with (97), it follows that

$$\mathscr{E}(u_k) - \mathscr{E}(u_k - s_{k,\mu} g(u_k)) \geqslant s_{k,\mu} \Big(1 - \frac{1}{2} s_{k,\mu} L\Big) \|g(u_k)\|^2. \tag{101}$$

Now, let $\mu_k$ denote the smallest $\mu = 0, 1, 2, \ldots$ for which $1 - \frac{L}{2} s_{k,\mu} \geqslant \sigma$. Such $\mu_k$ exists since both parameters of the Armijo rule (84) satisfy $0 < f, \sigma < 1$. Therefore,

$$1 - \frac{L}{2} s_{k,\mu} < \sigma \quad \text{for all } \mu = 0, \ldots, \mu_k - 1, \text{ and } 1 - \frac{L}{2} s_{k,\mu_k} \geqslant \sigma. \tag{102}$$

By applying (101) to $\mu = \mu_k$ and using the last inequality from (102), it follows that

$$\mathscr{E}(u_k) - \mathscr{E}(u_k - \widehat{s}_k g(u_k)) \geqslant \sigma \widehat{s}_k \|g(u_k)\|^2, \qquad \widehat{s}_k := s_{k,\mu_k}. \tag{103}$$

A comparison of (103) with the Armijo rule (84) shows that $j \leqslant \mu_k$. Therefore, the actual stepsize $s_k$ of the gradient descent in (83) satisfies $s_k \geqslant \widehat{s}_k$, and hence,

$$\begin{aligned} \mathscr{E}(u_k) - \mathscr{E}(u_{k+1}) &= \mathscr{E}(u_k) - \mathscr{E}(u_k - s_k g(u_k)) \\ &\geqslant \sigma s_k \|g(u_k)\|^2 \geqslant \sigma \widehat{s}_k \|g(u_k)\|^2, \end{aligned} \tag{104}$$

where use is also made of (72). Now, the first inequality in (102), considered for $\mu = \mu_k - 1$, implies that $\sigma > 1 - \frac{L}{2} s_{k,\mu_k - 1} = 1 - \frac{L}{2f} \widehat{s}_k$, which is equivalent to

$$\widehat{s}_k > \frac{2f}{L}(1-\sigma). \tag{105}$$

By combining (105) with the last inequality in (104) and subtracting $\mathscr{E}(u_*)$ from both sides of the resulting inequality, it follows

that

$$\mathscr{E}(u_{k+1}) - \mathscr{E}(u_*) \leqslant \mathscr{E}(u_k) - \mathscr{E}(u_*) - \frac{2f}{L}\sigma(1-\sigma)\|g(u_k)\|^2. \qquad (106)$$

Application of the second inequality from (94) to $u = u_k$ yields $\|g(u_k)\|^2 \geqslant 2\ell(\mathscr{E}(u_k) - \mathscr{E}(u_*))$ whose combination with (106) leads to

$$\mathscr{E}(u_{k+1}) - \mathscr{E}(u_*) \leqslant r(\mathscr{E}(u_k) - \mathscr{E}(u_*)) \qquad (107)$$

for all $k$ large enough, where $r$ is given by (99). Since (107) is satisfied for all sufficiently large $k$, then $\mathscr{E}(u_k)$ indeed approaches $\mathscr{E}(u_*)$ at a geometric rate described by the first asymptotic relation in (98). The second relation in (98) follows from the first one in view of the first inequality in (94). ■

Note that not every local minimum $u_*$ satisfies (94) with a positive constant $\ell$. As mentioned before, the existence of such a constant $\ell > 0$ is closely related to (93). "Degenerate" local minima (for which such $\ell$ does not exist) occur, for example, in the case when the transfer function of the corresponding coherent quantum controller is of McMillan degree less than $n$; see, for example, [8] and references therein.

## 12 Numerical Examples of Optimal CQLQG Controller Design

We will now present three examples which illustrate performance of the proposed numerical algorithm for the coherent quantum controller design.

**Example 8** *The gradient descent algorithm of Section 10 was tested to find a locally optimal solution of the CQLQG control problem for a PR quantum plant in (1) with dimensions $n = m_2 = p_1 = p_2 = r = 2$, $m_1 = 4$. The following state-space matrices $A$, $B$, $C$, $E$ of the plant were randomly generated subject to the PR conditions (19), (22), together and the weighting matrices $F$, $G$ in (8) (also provided are the standard feedthrough matrices $D$ and $d$):*

$$A = \begin{bmatrix} 0.9534 & -1.1165 \\ 0.4193 & 1.8821 \end{bmatrix}, \quad B = \begin{bmatrix} -1.7174 & -0.2189 & 1.9180 & 0.5636 \\ -0.6815 & 1.3570 & 0.2985 & -0.3679 \end{bmatrix},$$

$$C = \begin{bmatrix} -1.3570 & -0.2189 \\ -0.6815 & 1.7174 \end{bmatrix}, \quad D = \begin{bmatrix} 1 & 0 & 0 & 0 \\ 0 & 1 & 0 & 0 \end{bmatrix}, \quad d = \begin{bmatrix} 1 & 0 \\ 0 & 1 \end{bmatrix},$$

$$E = \begin{bmatrix} -0.3238 & 0.2779 \\ -1.1693 & -0.5966 \end{bmatrix}, \quad F = \begin{bmatrix} -0.8290 & -0.9665 \\ -1.8655 & -0.0357 \end{bmatrix}, \quad G = \begin{bmatrix} -0.2324 & -0.1608 \\ -0.5822 & -1.0961 \end{bmatrix}.$$

*This plant is unstable (the eigenvalues of the matrix $A$ are $1.4177 \pm 0.5025i$). The algorithm was run with parameters $h_{\max} = 1$, $f = 0.5$, $\sigma = 0.9$, $\varepsilon = 10^{-6}$ in (81)–(85) for 10 randomly generated stabilizing PR controllers as initial points. Starting from these points, it took 307 to 2318 steps for the algorithm to reach the fulfilment of the termination condition, with the average number of iterations being 1075. The local minimum value of the LQG cost is $\mathscr{E}_{\min} = 12.1026$ and is achieved at the following controller parameters:*

$$R = \begin{bmatrix} -0.5611 & -1.5567 \\ -1.5567 & 1.8283 \end{bmatrix}, \quad b = \begin{bmatrix} 1.8111 & 0.7201 \\ -1.4979 & -3.9696 \end{bmatrix}, \quad e = \begin{bmatrix} -0.1250 & 4.9673 \\ -4.4929 & -1.3387 \end{bmatrix}.$$

*The values of the LQG cost $\mathscr{E}(u_k)$ for the gradient descent sequences $u_k$ are presented in Fig. 2 in the form of semi-logarithmic graphs of $\frac{\mathscr{E}(u_k)}{\mathscr{E}_{\min}} - 1$. These graphs are in qualitative agreement with the relatively slow linear convergence rate, typical for gradient descent methods. However, they also show that the proposed algorithm is fairly reliable, being able to cope with poor initial approximations where the LQG cost $\mathscr{E}(u_0)$ exceeds the minimum value $\mathscr{E}_{\min}$ by an order of magnitude.*

In the next example, the gradient descent algorithm of Section 10 is used in order to solve the CQLQG control problem for a higher dimensional quantum plant.

**Example 9** *A PR quantum plant (1) with dimensions $n = m_2 = p_1 = p_2 = r = 4$, $m_1 = 8$ was produced with the following randomly generated state-space matrices $A$, $B$, $C$, $E$ (satisfying the PR conditions (19), (22)) together with the weighting matrices $F$, $G$ in (8) (and standard feedthrough matrices $D$ and $d$):*

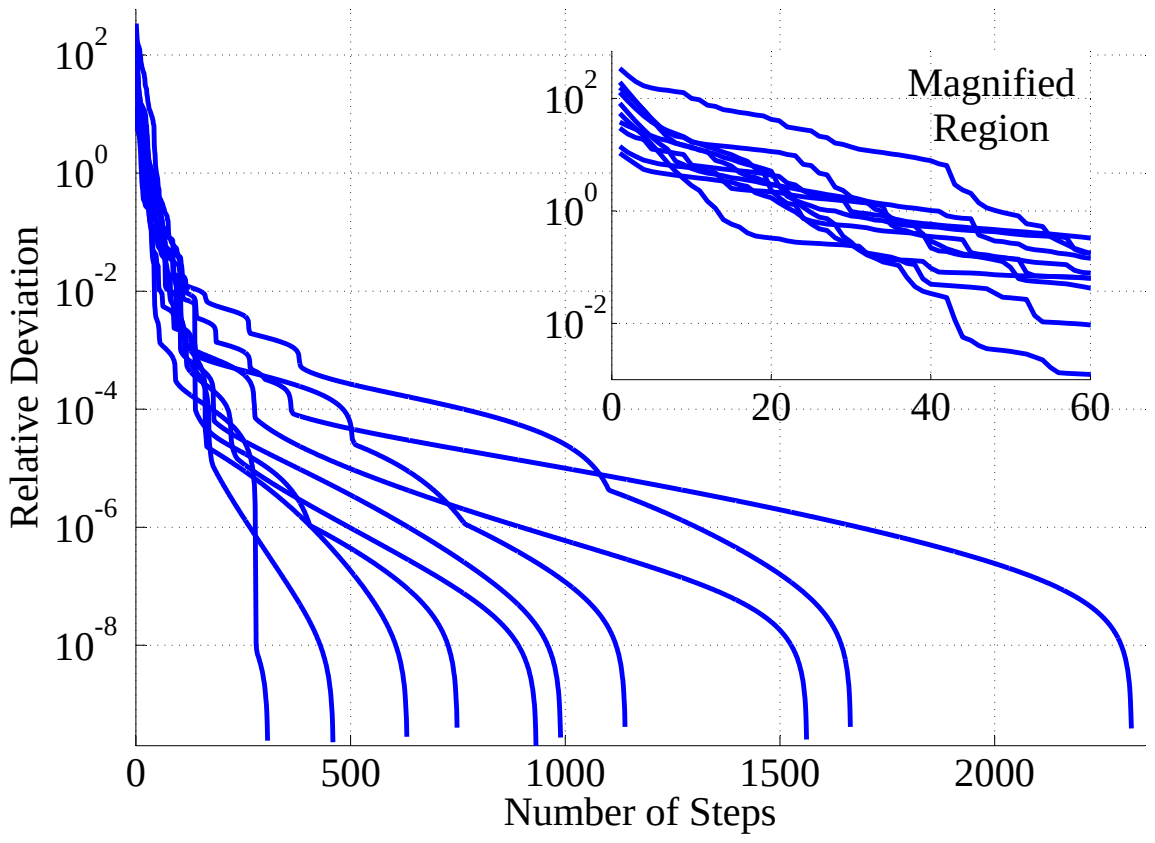

Fig. 2. The relative deviations $\frac{\mathscr{E}(u_k)}{\mathscr{E}_{\min}} - 1$ from the minimum value of the LQG cost on a logarithmic scale versus the number of steps $k$.

$$A = \begin{bmatrix} -1.4350 & 1.8015 & 0.6224 & -1.9677 \\ 4.2246 & 1.0737 & 0.0050 & 0.7269 \\ -2.2170 & -1.2044 & 0.8288 & 0.8162 \\ 0.3659 & 0.0671 & -1.8571 & 0.0519 \end{bmatrix}, \qquad B_1 = \begin{bmatrix} 0.2260 & 0.7044 & -1.2687 & -0.6255 \\ -0.2949 & 0.7202 & -1.3567 & -0.0782 \\ 1.1299 & 0.2743 & 0.5416 & -0.0526 \\ 1.7793 & 0.5586 & 0.6810 & 0.0559 \end{bmatrix},$$

$$B_2 = \begin{bmatrix} -1.9870 & 1.0095 & -0.6543 & -0.4582 \\ -2.8180 & 1.2571 & 1.9406 & 0.3029 \\ 0.2516 & -0.1793 & -0.4433 & -0.9985 \\ -0.6919 & -0.3315 & -0.8624 & 0.2582 \end{bmatrix}, \qquad C = \begin{bmatrix} -0.8595 & 0.6913 & 0.8501 & -0.8178 \\ -0.3597 & 1.2905 & -0.6205 & 0.5116 \\ -2.1776 & 0.4383 & 1.1181 & -0.4056 \\ -2.1282 & 0.3051 & -0.3811 & -0.0535 \end{bmatrix},$$

$$D = \begin{bmatrix} 1 & 0 & 0 & 0 & 0 & 0 & 0 & 0 \\ 0 & 1 & 0 & 0 & 0 & 0 & 0 & 0 \\ 0 & 0 & 1 & 0 & 0 & 0 & 0 & 0 \\ 0 & 0 & 0 & 1 & 0 & 0 & 0 & 0 \end{bmatrix}, \qquad d = I_4, \qquad E = \begin{bmatrix} 0.2872 & 0.6462 & -0.1964 & -0.5563 \\ -0.3272 & 0.0165 & 0.6345 & -1.1129 \\ 1.0013 & -0.7923 & 0.1927 & -0.8871 \\ -0.3072 & 0.1521 & -0.4705 & -1.2071 \end{bmatrix},$$

$$F = \begin{bmatrix} 0.3486 & -0.2299 & 1.2853 & 1.1289 \\ 1.6087 & -0.2542 & 0.4107 & 0.4351 \\ 1.0912 & 0.5867 & -0.9117 & 1.0435 \\ -1.5572 & 3.2263 & -0.4087 & -1.4700 \end{bmatrix}, \qquad G = \begin{bmatrix} 0.8558 & 1.9594 & -1.6661 & -0.7205 \\ -1.3948 & -1.1263 & 0.9523 & 0.2601 \\ -0.2111 & -1.8285 & -0.1029 & -0.9782 \\ 0.3339 & -0.8670 & -0.4206 & 0.6536 \end{bmatrix},$$

*where* $B := \begin{bmatrix} B_1 & B_2 \end{bmatrix}$. *This plant is unstable (the eigenvalues of the matrix* $A$ *are* $1.2215 \pm 1.1623i$, $-3.1804$, $1.2568$*). The algorithm was run with the parameters* $h_{\max} = 1$, $f = 0.333$, $\sigma = 0.9$, $\varepsilon = 10^{-6}$ *in (81)–(85) for three randomly generated stabilizing PR controllers as initial points in a neighbourhood of a local minimum which was found in advance. The local minimum value of the LQG cost is* $\mathscr{E}_{\min} = 274.0419$ *and is achieved at the following controller parameters:*

$$R = \begin{bmatrix} -3.0592 & -8.3578 & -2.1617 & -6.4581 \\ -8.3578 & 20.8551 & -2.6142 & 12.1693 \\ -2.1617 & -2.6142 & 10.2172 & -4.5633 \\ -6.4581 & 12.1693 & -4.5633 & 5.8481 \end{bmatrix},$$

$$b = \begin{bmatrix} 0.4141 & -2.4157 & -1.7490 & -2.8981 \\ -0.2109 & -0.0412 & 2.3931 & 1.3569 \\ 0.4802 & -1.6943 & -4.0817 & -3.5863 \\ 0.3551 & -0.1835 & -3.3398 & -2.3062 \end{bmatrix},$$

$$e = \begin{bmatrix} -14.3579 & -12.2259 & -25.2370 & 2.1155 \\ -5.1100 & 9.9388 & -12.8367 & -4.9427 \\ 1.0033 & -13.8569 & -1.6189 & 3.9519 \\ 13.7002 & -1.8789 & 10.8927 & 0.9094 \end{bmatrix}.$$

*We estimated the values $\ell = 2.3807 \times 10^{-7}$ and $L = 3.8940 \times 10^{-5}$ as the extreme eigenvalues of the Hessian operator of the LQG cost on the normal subspace at the local minimum. In view of Theorem 7, this ensures a linear convergence rate with $r = 0.9993$. The values of the LQG cost $\mathscr{E}(u_k)$ for the gradient descent sequences $u_k$ are presented as semi-logarithmic graphs of $\frac{\mathscr{E}(u_k)}{\mathscr{E}_{\min}} - 1$ in Fig. 3. These graphs are also in qualitative agreement with the linear convergence rate. They also show that the*

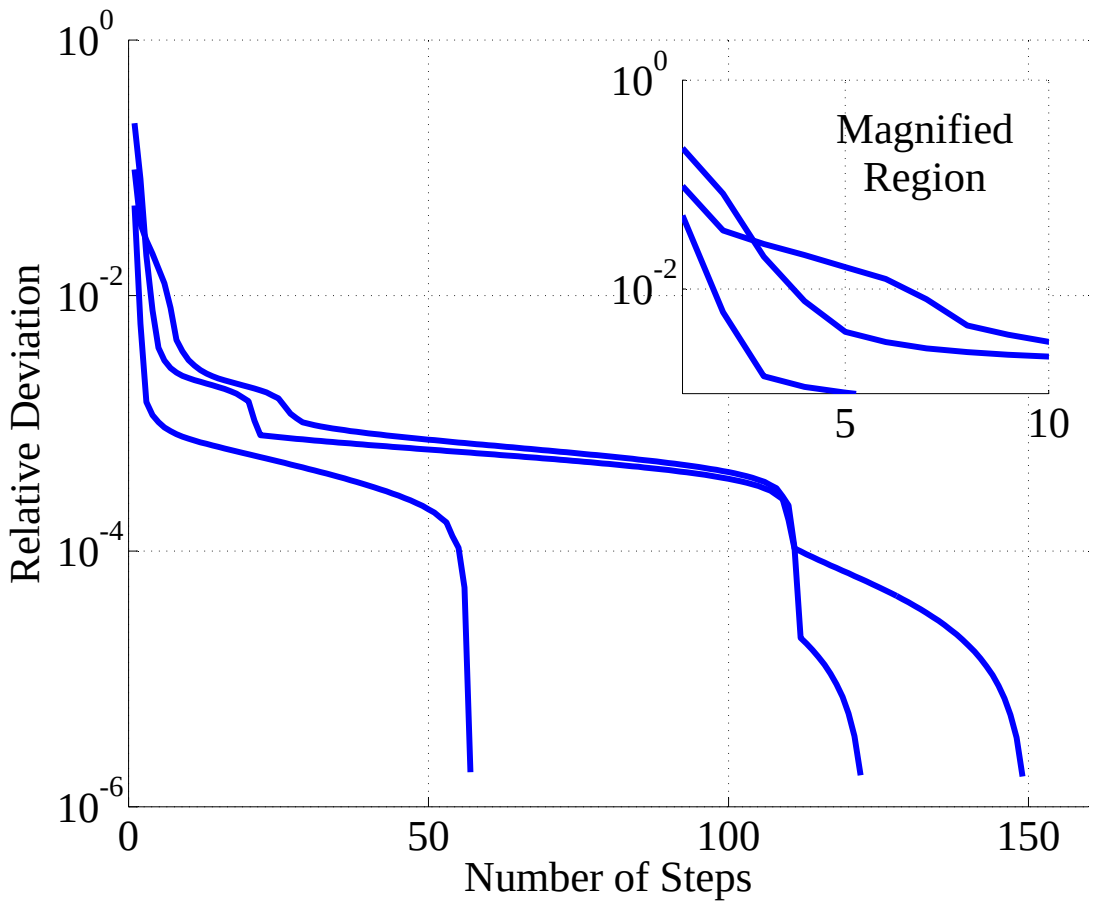

Fig. 3. The relative deviations $\frac{\mathscr{E}(u_k)}{\mathscr{E}_{\min}} - 1$ from the minimum value of the LQG cost on a logarithmic scale versus the number of steps $k$.

*proposed algorithm is fairly reliable in the higher dimensional case considered.*

The next example shows that the numerical algorithm can be used in control synthesis problems for practical quantum optical systems.

**Example 10** *Following the example of [21, Section 8], consider the dynamics of an atom trapped in an optical cavity. This setup is modelled as a PR quantum plant governed by (1) with the following state-space matrices A, B, C, D, E and the* *weighting matrices F, G in (8):*

$$A = \begin{bmatrix} 0 & \Delta \\ -\Delta & 0 \end{bmatrix}, \qquad B = \begin{bmatrix} 0 & 0 & 0 & 0 \\ 0 & -2\sqrt{k_1} & 0 & -2\sqrt{k_2} \end{bmatrix},$$
$$C = \begin{bmatrix} 2\sqrt{k_1} & 0 \\ 0 & 0 \end{bmatrix}, \quad D = \begin{bmatrix} 1 & 0 & 0 & 0 \\ 0 & 1 & 0 & 0 \end{bmatrix}, \quad d = \begin{bmatrix} 1 & 0 \\ 0 & 1 \end{bmatrix},$$
$$E = \begin{bmatrix} 0 & 0 \\ 0 & -2\sqrt{k_3} \end{bmatrix}, \quad F = \begin{bmatrix} 1 & 0 \\ 0 & 1 \end{bmatrix}, \qquad G = \begin{bmatrix} 1 & 0 \\ 0 & 1 \end{bmatrix},$$

*where the parameters* $\Delta = 0.1$ *and* $k_1 = k_2 = k_3 = 0.01$ *are the same as in the above mentioned example of [21]. This plant is marginally stable (the eigenvalues of the matrix A are* $\pm\Delta i$*). The algorithm in (81)–(85) was run with the parameters* $h_{\max} = 1$*,* $f = 0.5$*,* $\sigma = 0.9$*,* $\varepsilon = 10^{-6}$ *for ten randomly generated stabilizing PR controllers as initial points. The local minimum value of the LQG cost is* $\mathscr{E}_{\min} = 2.0418$ *and is achieved at the following controller parameters:*

$$R = \begin{bmatrix} -0.2488 & 0.0900 \\ 0.0900 & -0.0395 \end{bmatrix}, \quad b = \begin{bmatrix} -0.1403 & -0.2305 \\ -0.0297 & -0.8185 \end{bmatrix}, \quad e = \begin{bmatrix} -0.3185 & -0.0785 \\ 0.1940 & -0.3017 \end{bmatrix}. \qquad (108)$$

*The eigenvalues of the corresponding closed-loop system matrix* $\mathscr{A}$ *in (11) are* $-0.0245 \pm 0.1019i$ *and* $-0.0852 \pm 0.0485i$*. It took* 9356 *steps and* 29.45 *seconds on average for the algorithm to reach the fulfillment of the termination condition.* [1] *These results can be compared with [21, Example I], where the suboptimal algorithm achieved the value* $\mathscr{E}_{\min} = 5.7382$ *after* 1000 *iterations of a semi-definite program solver with the running time of* 2944.9 *seconds. Experimental implementation of the coherent quantum controller with the parameters (108) can be carried out using the network synthesis theory [20] for quantum optical systems.*

## 13 Conclusion

A numerical approach has been proposed for solving the optimal CQLQG controller design problem using the gradient descent method. We have taken into account the symmetry of the problem with respect to symplectic similarity transformations of coherent quantum controllers and introduced corresponding equivalence classes of state-space realizations of such controllers. We have proposed a modified gradient flow which is concerned with norm-balanced realizations of coherent quantum controllers. A numerical algorithm has been developed for finding a locally optimal solution of the CQLQG control problem and its convergence has been investigated. The algorithm has been tested and appeared to be fairly reliable in numerical experiments with randomly generated stabilizing PR controllers as initial approximations. The algorithm compares favourably in comparison with the previous results. The lack of a more systematic method for initialization and a relatively slow linear convergence rate (in comparison with Newton-like optimization methods) are the main shortcomings of the algorithm. These issues are a subject for future research and will be tackled in subsequent publications.

## Acknowledgements

This work is supported by the Australian Research Council.

[1] All computations were performed on MATLAB R2012b running on a HP Z220 SFF workstation with Intel Core i7-3770 CPU working at 3.4 GHz and with 8 GB of DDR3 RAM.

## A Computing the first-order Fréchet derivatives of the LQG cost in Lemma 1

The ALEs (18), (36) and the properties of the Frobenius inner product of matrices allow the LQG cost $\mathscr{E}$ in (27) to be represented as

$$\mathscr{E}(u) = \frac{1}{2}\langle \mathscr{C}^{\mathrm{T}}\mathscr{C}, P\rangle = \frac{1}{2}\langle Q, \mathscr{B}\mathscr{B}^{\mathrm{T}}\rangle = -\langle H, \mathscr{A}\rangle, \tag{A1}$$

where $H$ is the Hankelian given by (37). In what follows, $\delta(\cdot)$ denotes the first variation, and $\delta_X(\cdot)$ is the first variation with respect to an independent matrix-valued variable $X$. Since the matrix $R$ influences the LQG cost $\mathscr{E}$ in (A1) only through the controller matrix $a$ in (24) which enters the matrix $\mathscr{A}$ in (11), then

$$\begin{aligned}\delta_R\mathscr{E} &= \frac{1}{2}\langle \mathscr{C}^{\mathrm{T}}\mathscr{C}, \delta_R P\rangle \\ &= -\frac{1}{2}\langle \mathscr{A}^{\mathrm{T}}Q + Q\mathscr{A}, \delta_R P\rangle \\ &= -\frac{1}{2}\langle Q, \mathscr{A}\delta_R P + \delta_R P\mathscr{A}^{\mathrm{T}}\rangle \\ &= \frac{1}{2}\langle Q, \delta_R\mathscr{A}P + P\delta_R\mathscr{A}^{\mathrm{T}}\rangle \\ &= \langle H, \delta_R\mathscr{A}\rangle = \langle H_{22}, \delta_R a\rangle = \langle H_{22}, 2\Theta_2\delta R\rangle.\end{aligned} \tag{A2}$$

Now, since $R\in\mathbb{S}_n$, and the subspaces $\mathbb{S}_n$ and $\mathbb{A}_n$ are orthogonal, then the first variation of the LQG cost in (A2) takes the form $\delta_R\mathscr{E} = 2\langle\Theta_2^{\mathrm{T}}H_{22}, \delta R\rangle = -2\langle \mathrm{sym}(\Theta_2 H_{22}), \delta R\rangle$, which, by the definition of the Fréchet derivative, establishes (38). By a similar reasoning, the first variations of the LQG cost with respect to $b$ and $e$ are as follows:

$$\begin{aligned}\delta_b\mathscr{E} =& \langle H, \delta_b\mathscr{A}\rangle + \langle Q\mathscr{B}, \delta\mathscr{B}\rangle + \langle \mathscr{C}P, \delta_b\mathscr{C}\rangle \\ =& \langle H_{22}, \delta_b a\rangle + \langle E^{\mathrm{T}}H_{12}, \delta_b c\rangle + \langle (Q\mathscr{B})_{22}, \delta b\rangle \\ &+ \langle G^{\mathrm{T}}FP_{12} + G^{\mathrm{T}}GcP_{22}, \delta_b c\rangle \\ =& \langle -\mathrm{asym}(H_{22}\Theta_2^{-1})bJ_2 + Q_{21}Ed + Q_{22}b \\ &- \Theta_2^{-1}(H_{12}^{\mathrm{T}}E + P_{21}F^{\mathrm{T}}G + P_{22}c^{\mathrm{T}}G^{\mathrm{T}}G)dJ_2, \delta b\rangle, \\ \delta_e\mathscr{E} =& \langle H_{21}, \delta e\, C\rangle + \langle H_{22}, \delta_e a\rangle + \frac{1}{2}\langle Q, \delta_e(\mathscr{B}\mathscr{B}^{\mathrm{T}})\rangle \\ =& \langle H_{21}C^{\mathrm{T}}, \delta e\rangle + \langle H_{22}\Theta^{-1}, \mathrm{asym}(\delta eDJ_1D^{\mathrm{T}}e^{\mathrm{T}})\rangle \\ &+ \langle Q\mathscr{B}, \delta_e\mathscr{B}\rangle \\ =& \langle H_{21}C^{\mathrm{T}} - \mathrm{asym}(H_{22}\Theta^{-1})eDJ_1D^{\mathrm{T}} \\ &+ (Q_{21}B + Q_{22}eD)D^{\mathrm{T}}, \delta e\rangle,\end{aligned}$$

which leads to (39) and (40) in view of (26), (41) and (42).

## B Computing the second-order Gâteaux derivative of the LQG cost

The first-order Gâteaux derivative of the LQG cost $\mathscr{E}$ along the gradient $g$ in (33) is expressed in terms of the first-order Fréchet derivatives from (38)–(40) by using (75) and (76), provided $u\in\mathbb{U}_0$. Hence, the second-order Gâteaux derivative along the gradient can be computed as

$$\begin{aligned}\mathscr{D}_g^2\mathscr{E} &= \langle \partial_u^2\mathscr{E}(u)(g), g\rangle = \langle \mathscr{D}_g g, g\rangle \\ &= \langle \mathscr{D}_g\partial_R\mathscr{E}, \partial_R\mathscr{E}\rangle + \langle \mathscr{D}_g\partial_b\mathscr{E}, \partial_b\mathscr{E}\rangle + \langle \mathscr{D}_g\partial_e\mathscr{E}, \partial_e\mathscr{E}\rangle,\end{aligned} \tag{B1}$$

where

$$\mathscr{D}_g\partial_R\mathscr{E} = -2\mathrm{sym}(\Theta_2\mathscr{D}_g H_{22}), \tag{B2}$$

$$\begin{aligned}\mathscr{D}_g\partial_b\mathscr{E} =& \mathscr{D}_g Q_{21}Ed + \mathscr{D}_g(Q_{22}b) \\ &- \mathscr{D}_g(\psi b)J_2 - \mathscr{D}_g\chi dJ_2,\end{aligned} \tag{B3}$$

$$\begin{aligned}\mathscr{D}_g\partial_e\mathscr{E} =& \mathscr{D}_g H_{21}C^{\mathrm{T}} + \mathscr{D}_g Q_{21}BD^{\mathrm{T}} \\ &+ \mathscr{D}_g(Q_{22}e) - \mathscr{D}_g(\psi e)DJ_1D^{\mathrm{T}}.\end{aligned} \tag{B4}$$

The second-order Gâteaux derivative in (B1) can now be computed by using (B2)–(B4), the Leibniz product rule and the first-order Gâteaux derivatives of $P$, $Q$. To this end, by differentiating both sides of the ALE (18) and its dual (36), it follows that the matrices $\mathscr{D}_g P$ and $\mathscr{D}_g Q$ are unique solutions of the ALEs

$$\mathscr{A}\mathscr{D}_g P + \mathscr{D}_g P \mathscr{A}^{\mathrm{T}} + 2\mathrm{sym}(\mathscr{D}_g \mathscr{A} P + \mathscr{D}_g \mathscr{B}\mathscr{B}^{\mathrm{T}}) = 0,$$
$$\mathscr{A}^{\mathrm{T}}\mathscr{D}_g Q + \mathscr{D}_g Q \mathscr{A} + 2\mathrm{sym}(\mathscr{A}^{\mathrm{T}}\mathscr{D}_g Q + \mathscr{C}^{\mathrm{T}}\mathscr{D}_g \mathscr{C}) = 0.$$

Here, in view of (11), (24), (25) and the relation $\mathscr{D}_g u = g$ (whose component-wise form is $\mathscr{D}_g R = \partial_R \mathscr{E}$, $\mathscr{D}_g b = \partial_b \mathscr{E}$ and $\mathscr{D}_g e = \partial_e \mathscr{E}$ in accordance with (33)),

$$\mathscr{D}_g \mathscr{A} = \begin{bmatrix} 0 & -EdJ_2 \partial_b \mathscr{E}^{\mathrm{T}} \Theta_2^{-1} \\ \partial_e \mathscr{E} C & \mathscr{D}_g a \end{bmatrix},$$
$$\mathscr{D}_g \mathscr{B} = \begin{bmatrix} 0 & 0 \\ \partial_e \mathscr{E} D & \partial_b \mathscr{E} \end{bmatrix},$$
$$\mathscr{D}_g \mathscr{C} = \begin{bmatrix} 0 & -GdJ_2 \partial_b \mathscr{E}^{\mathrm{T}} \Theta_2^{-1} \end{bmatrix},$$

with

$$\mathscr{D}_g a = 2\Theta_2 \partial_R \mathscr{E} - \mathrm{asym}(\partial_e \mathscr{E} D J_1 D^{\mathrm{T}} e^{\mathrm{T}} + \partial_b \mathscr{E} J_2 b^{\mathrm{T}})\Theta_2^{-1}.$$